# Solar Active Regions Detection Via 2D Circular Kernel Time Series Transformation, Entropy and Machine Learning Approach

Irewola Aaron Oludehinwa[1,2,*], Andrei Velichko[3], Maksim Belyaev[3], and Olasunkanmi I. Olusola[2]

1. Department of Physics, Caleb University, Lagos, Nigeria
2. Department of Physics, University of Lagos, Lagos, Nigeria
3. Institute of Physics and Technology, Petrozavodsk State University, 185910 Petrozavodsk, Russia

***Correspondence:** irewola.oludehinwa@calebuniversity.edu.ng; Tel.: +2348068030109

**Abstract**

This study proposes an enhancement to the existing method for detecting Solar Active Regions (ARs). Our technique tracks ARs using images from the Atmospheric Imaging Assembly (AIA) of NASA's Solar Dynamics Observatory (SDO). It involves a 2D circular kernel time series transformation, combined with Statistical and Entropy measures, and a Machine Learning (ML) approach. The technique transforms the circular area around pixels in the SDO AIA images into one-dimensional time series (1-DTS). Statistical measures (Median Value, $X_{med}$; 95th Percentile, $X_{95}$) and Entropy measures (Distribution Entropy, DisEn; Fuzzy Entropy, FuzzyEn) are used as feature selection methods (FSM 1), alongside a method applying 1-DTS elements directly as features (FSM 2). The ML algorithm classifies these series into three categories: no Active Region (nARs type 1, class 1), non-flaring Regions outside active regions with brightness (nARs type 2, class 2), and flaring Active Regions (ARs, class 3). The ML model achieves a classification accuracy of 0.900 and 0.914 for Entropy and Statistical measures, respectively. Notably, Fuzzy Entropy shows the highest classification accuracy ($A_{KF}$=0.895), surpassing DisEn ($A_{KF}$=0.738), $X_{95}$ ($A_{KF}$=0.873), and $X_{med}$ (AKF=0.840). This indicates the high effectiveness of Entropy and Statistical measures for AR detection in SDO AIA images. FSM 2 captures a similar distribution of flaring AR activities as FSM 1. Additionally, we introduce a generalizing characteristic of AR activities (GSA), finding a direct agreement between increased AR activities and higher GSA values. The Python code implementation of the proposed method is available in supplementary material.

## 1. Introduction

The Sun has a major impact on Earth: it provides the light and energy that are vital to life on our planet and greatly shapes the Earth's climate. However, the Sun's activity evolves and its dynamics can be in the state of quiet or disturbed. Its disturbances are associated with an intense localized eruption of plasma which are obviously seen in Active Regions (ARs) of Atmospheric Imaging Assembly (AIA) of Solar Dynamics Observatory (SDO) as solar flares with a well-defined area comprising strong magnetic fields. They are sporadic brightening observed over the Sun's surface and usually occur when an accumulated magnetic energy is released into the solar atmosphere [1–6]. The erupted plasmas are accompanied by coronal mass ejection, solar particle events, and other solar phenomenon that are potentially harmful to spacecraft technology and astronauts in space, as well to terrestrial infrastructure, including our power grid, telecommunication systems, and radio operations. This sporadic eruption of plasmas over the Sun's surface highlights the importance of understanding the dynamical evolution of the Sun's activity and its associated space-weather conditions in the heliosphere.

The imagery from solar observatories is one of the most valuable sources of information regarding the activity of the Sun. As a result, Solar Dynamics Observatory (SDO) mission of the National Aeronautics Space Agency (NASA) captures approximately 70,000 images of the Sun activity in a day [7–10]. Notably, the continuous visual inspection of these images of solar observatory regarding the Solar activity is challenging. There is a need to develop an innovative technique that can automatically detect and track the ARs in the Sun images for a more defined and robust search of the space weather. Different methods of classifying, detecting, and capturing the activity of the Sun have been proposed by several authors using Spectrogram, Image processing, Deep learning, and Machine Learning [2, 7, 19–21, 11–18]. For instance, in the work of [22] the authors developed an approach named Solar Demon to provide fast detections of flares, dimming, and EUV waves for space weather purposes. The automatic detections of Solar Demon enable them to build large catalogs with quantifiable results on a reproducible basis. [23] proposed a novel approach combining machine learning (ML) with feature selection to predict solar flares using Magnetic Features (MFs) from the Solar Monitor Active Region Tracker (SMART). Their algorithm correlates MFs with flares to identify flaring and non-flaring patterns, and applies ML and feature selection techniques to improve prediction accuracy. The approach outperformed traditional ML-based methods, such as the Automated Solar Activity Prediction (ASAP) system, in a comparative evaluation using standard forecast verification measures. [24] used the Support Vector Machine (SVM) algorithm to predict M- and X-class flares. They analyzed four years of data from the Helioseismic and Magnetic Imager (HMI) on the Solar Dynamics Observatory (SDO), creating a catalog of 2071 active regions, with 25 parameters describing each. Their method achieved high True Skill Statistics (TSS) scores, demonstrating strong predictive capabilities for distinguishing flaring and non-flaring active regions. In another study, [25] used machine learning to predict large solar flares from magnetograms captured by the Helioseismic and Magnetic Imager on SDO. They converted magnetic field data into Zernike moments (ZMs) and fed them into a Support Vector Machine (SVM) classifier, achieving accurate predictions up to 48 hours in advance. Out of 564 magnetograms, they correctly predicted 375 flaring regions, with only 10 false negatives and 21 false positives. [26] developed a solar flare prediction system using deep learning on 1-minute GOES X-ray flux data. The system consists of three neural networks: the first converts flux data into Markov Transition Field (MTF) images, the second extracts features from MTF images using unsupervised learning, and the third uses a Deep Convolutional Neural Network (CNN) to generate predictions from the learned features and MTF images. The system showed promising results, and the authors suggested that further improvements in accuracy can be achieved by leveraging advanced machine learning classification capabilities. [27] introduced an algorithm to automatically extract features from 5.5TB of SDO image data, covering the solar photosphere, chromosphere, transition region, and corona. The algorithm combines these features with historical flare data and physical process insights to predict solar flares within 2-24 hours. Optimizing for True Skill Score (TSS), they found that combining photospheric vector magnetic field data with flaring history yields the best performance. [28] studied the predictive capabilities of magnetic features properties generated by the Solar Monitor Active Region Tracker (SMART). They use marginal relevance as a filter features selection method to identify the most useful SMART magnetic feature properties such as region size, total flux, flux imbalance, flux emergence rate, Schrijver's R-value and Falconer's measurement of non-potentiality for separating flaring from non-flaring ARs. They also applied the logistic regression to derive classification rules to predict future observation and obtained significantly better results, i.e. True Skill (TSS)=0.84. [29] tested the applicability of Flare Likelihood and Region Eruption Forecasting (FLARECAST) on regular photospheric magnetic field provided by the Helioseismic and Magnetic Imager (HMI) from SDO. They showed the efficiency of the technique as predictors of flaring activity on a representative sample of active regions. It was also found that Ising energy appears to be an efficient predictor. [30] created a solar flare prediction tool using Support Vector Machine (SVM) classification on Zernike moments extracted from Active Region (AR) images. They utilized data from SDO's HMI and AIA instruments. Their results showed a power-law relationship between the time an AR appears and the first major flare (X- or M-class), with most significant flares occurring within 150 hours. [31] conceived the idea that the properties of the Polarity Inversion Line (PIL) in solar active regions (ARs) are strongly correlated with flare occurrence. The study used an unsupervised machine learning algorithm named Kernel Principal Component Analysis (KPCA) to derive features from the PIL mask and difference in PIL mask. Those features were classified into two categories, namely, the non-strong flaring ARs and the strong flaring ARs. Their results showed that features derived from the PIL mask by KPCA are effective in predicting flare occurrence. Recently, [32] used a convolutional neural network (CNN) to analyze the solar observations obtained from the Atmospheric Imaging Assembly (AIA). They identified each image by classifying the shape and position of the flare ribbons into two-ribbon flare, compact/circular ribbon flare, limb flare and the state where flaring regions are not present. The authors concluded that the network created can classify flare ribbon observation into any of the four classes with 94% accuracy. More recently, [33] introduced the detection and EUV flare tracking tool that can identify flare signatures and their precursors using extreme-ultraviolet (EUV) solar observation. The authors reported that the tracking tool can identify the location of disturbances and distinguish events occurring at the same time in multiple locations. Further work by [34] applied an image processing technique to automatically detect active regions of the Sun. The image processing technique was based on image enhancement, segmentation, pattern recognition and mathematical morphology. They reported that the identification and classification of Sunspots are useful techniques for tracking and predicting the solar activity. [17] used faster-R-CNN (Region with Convolutional Neural Networks) and YOLOV3 (You Only Look Once, Version 3) to learn the characteristics of active regions and employed a deep learning-based detection model for active regions. It was recorded that the performance demonstrates high accuracy of active region detection. For the faster-R-CNN model for ARs detection, the True Positive (TP) rate is 90%, while the True Negative (TN) is 98%. Also, for the YOLO V3 model for ARs detection, the TP rate and TN rate is 94% and 99% respectively. [35] examined the chromospheric and coronal properties of solar active regions through Active Region Patches (AARPs). It was reported that the AARPs database enables physics-informed parameterization and analysis using nonparametric discriminant. [36] investigated whether coronal, transition region, and chromospheric emission parameters from SDO/AIA images could predict imminent solar flares. Analyzing a large sample of active region images, they found that moment analysis-based parameters from direct and running-difference images effectively distinguished flaring regions using non-parametric discriminant analysis, providing physically meaningful results.

Despite the implementation of several approaches of tracking the activity of the Sun from the AIA images obtained from solar observatory, the concept of capturing and transforming the areas of the Active Region (ARs) in the SDO AIA image into 1-Dimensional Time Series (1-DTS) has not been considered in the literature to the best of our knowledge. The concept has the computational advantage of capturing accurate information regarding the ARs in the solar observatory image. This observation forms the bedrock of this present study, to track the Sun activity by capturing the flaring ARs areas in the SDO AIA image. Therefore, a tracking method is developed for ARs detection in SDO AIA images by classifying ARs. In this method, we take the advantage of Machine Learning algorithms on 2D circular kernel time series and Entropy measures. Recall that Machine Learning methods can be computationally demanding, in that they required large amount of data to train that might not be readily available. However, the 2D circular kernel time series transformation of solar observatory images into 1-DTS addresses these challenges by automatically transforming the ARs areas in the AIA images into series for feature extraction in Machine Learning. This effort will contribute to the operational space weather forecast.

Statistical measures are popular features to identify the pattern and trend of a dataset as explained in [37]. Entropy measures are essential tool used to measure the irregularity or randomness of a time series data and serve as an index to measure the complexity of a dynamical system [38–40]. Entropy concept is developed from information theory and can be used as a potential tool in change detection and image quality assessment. Image processing through entropy measures have been proposed by [41–43] in remote sensory, where Entropy is considered as an irregularity measure for images. Therefore, Entropy as a nonlinear tool possesses the potential to serve as a feature extraction in machine learning classification for solar activity.

In this paper, we propose a method for automatic detection of solar active regions. Figure 1 illustrates the main steps for classifying the areas in the AIA images with flaring Active Regions (ARs) and no Active Regions (nARs). First, a 2D circular kernel with radius $R$ is built that transforms the surface area of ARs into a 1-Dimensional Time Series (1-DTS). The 1-DTS is subjected to Statistical and Entropy measures for Machine Learning (ML) classification into three classes namely: no Active Region (nARs, type 1-class 1), non-flaring Region outside active region with brightness (nARs, type 2-class 2) and flaring Active Regions (ARs-class 3) respectively.

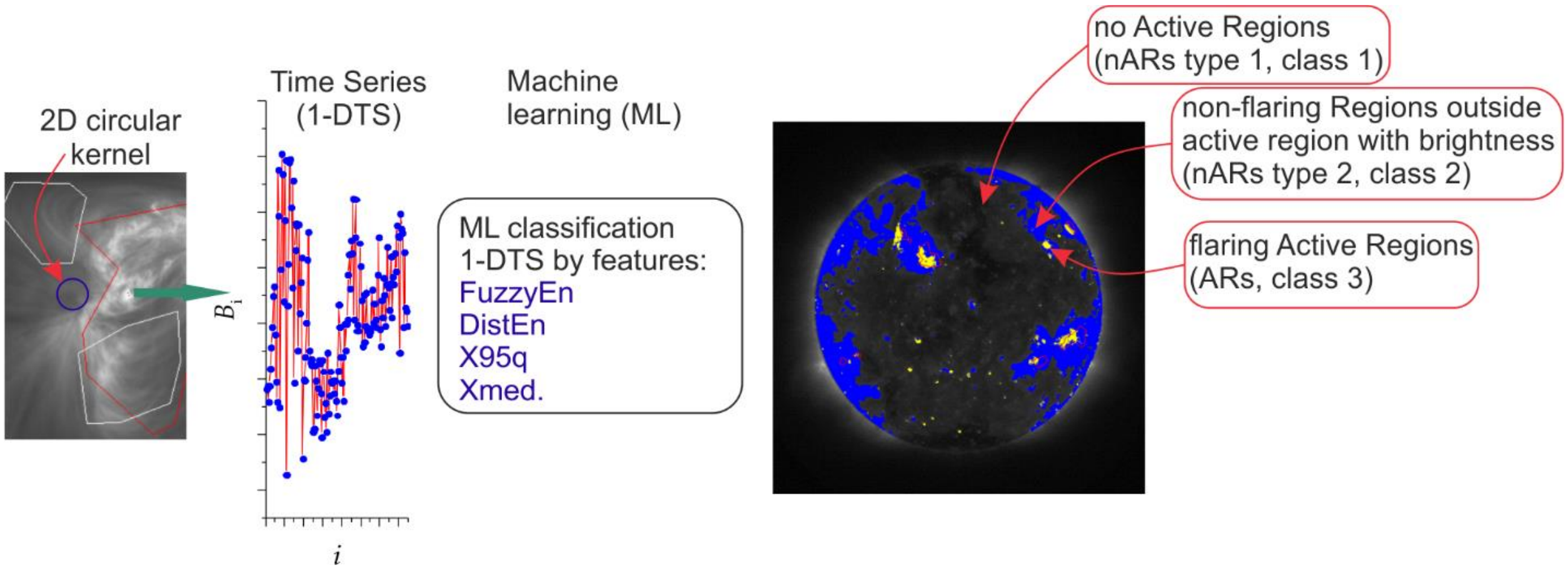

Figure 1: Schematic representation of the proposed classification method illustrated on an AIA image captured on June 5, 2021, utilizing a 2D circular kernel. The red color indicates the ARs contours obtained using HEK through SPoCA model. Also, a 2D circular kernel with radius $R$ = 14 pixels is built that transforms the surface area of ARs into a 1-Dimensional Time Series (1-DTS). The 1-DTS (Number of pixels, $N$ = 613) is subjected to statistical and Entropy features ($X_{95q}$, $X_{med}$ FuzzyEn, DistEn) for Machine Learning (ML) classification into three classes namely: flaring Active Regions (ARs), no Active Region (nARs type 1) and non-flaring Region outside active regions with brightness (nARs type 2) respectively. The blue color is the labelling of non-flaring Regions outside active region with brightness captured by ML classification (Class 2-nARs type 2). The class 2-nARs type 2 are also no active regions category while the yellow color is the labelling of the flaring ARs captured by ML classification (Class 3-ARs).

The major contributions of the paper are as follows:

- A new technique for detecting Solar Active Regions in the solar observatory images is developed by transforming the AIA image into One-Dimensional time series (1-DTS) using 2D circular kernel, Entropy measures with Machine learning approach.
- Entropy measure shows high potency as a feature extraction in capturing the ARs activities in the solar observatory images.
- A new method for estimating the generalized characteristics of ARs activities (GSA) captured from the pixels of the AIA images is developed.
- The idea of capturing flaring Active Regions and non-flaring Regions for machine learning classification is introduced. An implementation of the algorithms in Python is presented.

The rest of the paper is organized as follows. In section 2, the image dataset acquisition, description of the image data, the method of 2D circular kernel time series transformation, Statistical and Entropy measures used for training a machine learning algorithm are explained in detail. We present the results in section 3, followed by discussion in section 4. The conclusion is drawn in section 5.

## 2. Materials and Methods

A simple schematic for detecting Active Regions in SDO AIA images is outlined in Figure 1. The process for analyzing each AIA image comprises the following steps:

1) Load an AIA image, in its original size (4096 × 4096 pixels), and format (FITS).

2) Identify base pixels within the image spaced 8 pixels apart. These pixels form a 512x512 pixel matrix.

3) Surround each base pixel with a circle of radius R=14, termed a 2D Circular Kernel. Each kernel is transformed into a one-dimensional time series (1-DTS), resulting in a total of 262,144 time series.

4) Analyze each time series using Statistical and Entropy features ($X_{95q}$, $X_{med}$, FuzzyEn, DistEn), with each series containing only four features.

5) Use a supervised machine learning algorithm to classify regions, using Statistical and Entropy features, into three categories, namely: no Active Regions (nARs type 1), non-flaring Region outside active region with brightness (nAR type 2), and flaring Active Region (ARs).

6) Consequently, each base pixel within the circle is assigned a specific class and displayed in a corresponding color to illustrate the distribution of solar activity classes within the AIA image. A python package of our method for detecting Solar Active Regions in the AIA images of SDO is available in the supplementary material section.

A detailed exposition of each phase of the technique, including image acquisition, circular kernel extraction, and machine learning methodologies, is provided in the following subsections.

*2.1 Description of Image Dataset from Solar Dynamics Observatory*

To find Active Regions in the SDO AIA images, the Heliophysics Knowledge Base, HEK, [44–46] was accessed using the SunPy-library [47]. **In order** to detect Active Regions, HEK uses Spatial Possibilistic Clustering Algorithm, SPoCA, [2]. For this purpose, images obtained using the SDO AIA with wavelengths of 171Å and 193Å are used. Image processing is performed using Histogram-based Possibilities C-means (HPCM2) classifier [2] using the median of the last 10 computed class centers, followed by morphological operations on the pixels of the ARs class to group them into regions. Small regions with an area less than 1500 arcsec$^2$ are discarded. The data obtained from HEK represents the coordinates of the AR contours and timestamp. Then, using the SunPy-library and the Joint Science Operations Center database, JSOC, [48], an image with the nearest timestamp obtained using AIA in FITS format from JSOC keywords for metadata [49] at Level 1, 12 second cadence of wavelength 193 Å was acquired. The list of the selected solar flare events used in this study is shown in (Table 1). The retrieved AR contours are superimposed on this image based on the helioprojective-cartesian coordinate grid.

**Table 1:** List of selected solar flare events examined in this study.

| S/N | Solar flares Event date | Time of |
|---|---|---|

| | | Occurrence (UTC) | Solar Flare Class |
|---|---|---|---|
| Data 1 | 2021/06/05 | 20:01 | B |
| Data 2 | 2021/06/24 | 19:27 | B |
| Data 3 | 2021/07/09 | 14:42 | C |
| Data 4 | 2021/09/05 | 18:00 | B |
| Data 5 | 2021/12/19 | 09:16 | C |
| Data 6 | 2021/09/24 | 09:01 | C |
| **Data 7** | **2024/05/06** | **06:38** | **X** |
| **Data 8** | **2024/05/09** | **09:13** | **X** |
| **Data 9** | **2024/05/13** | **09:44** | **M** |
| **Data 10** | **2024/06/08** | **01:49** | **M** |

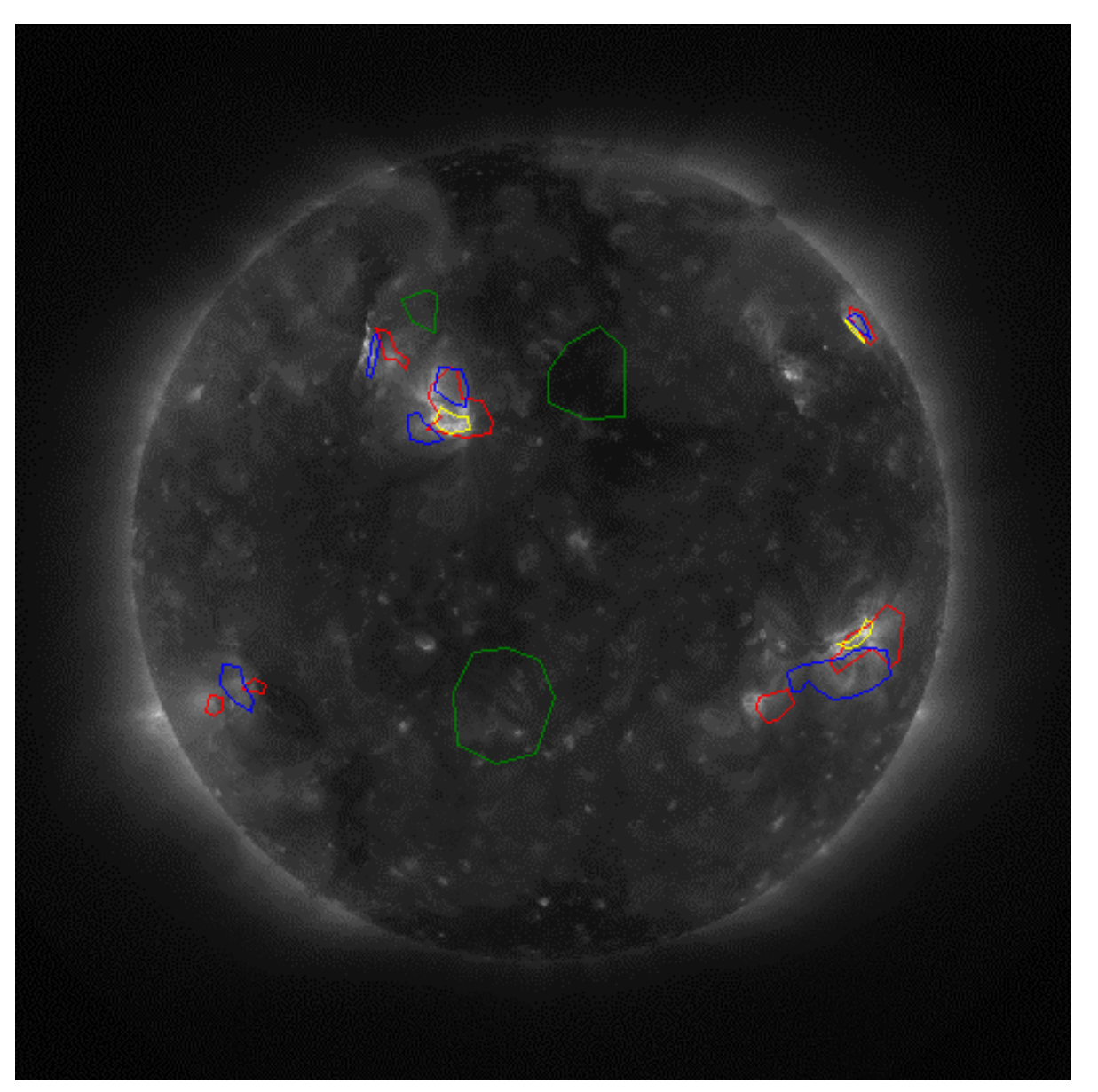

(a) Data 1 (2021/06/05) at 20:01UT

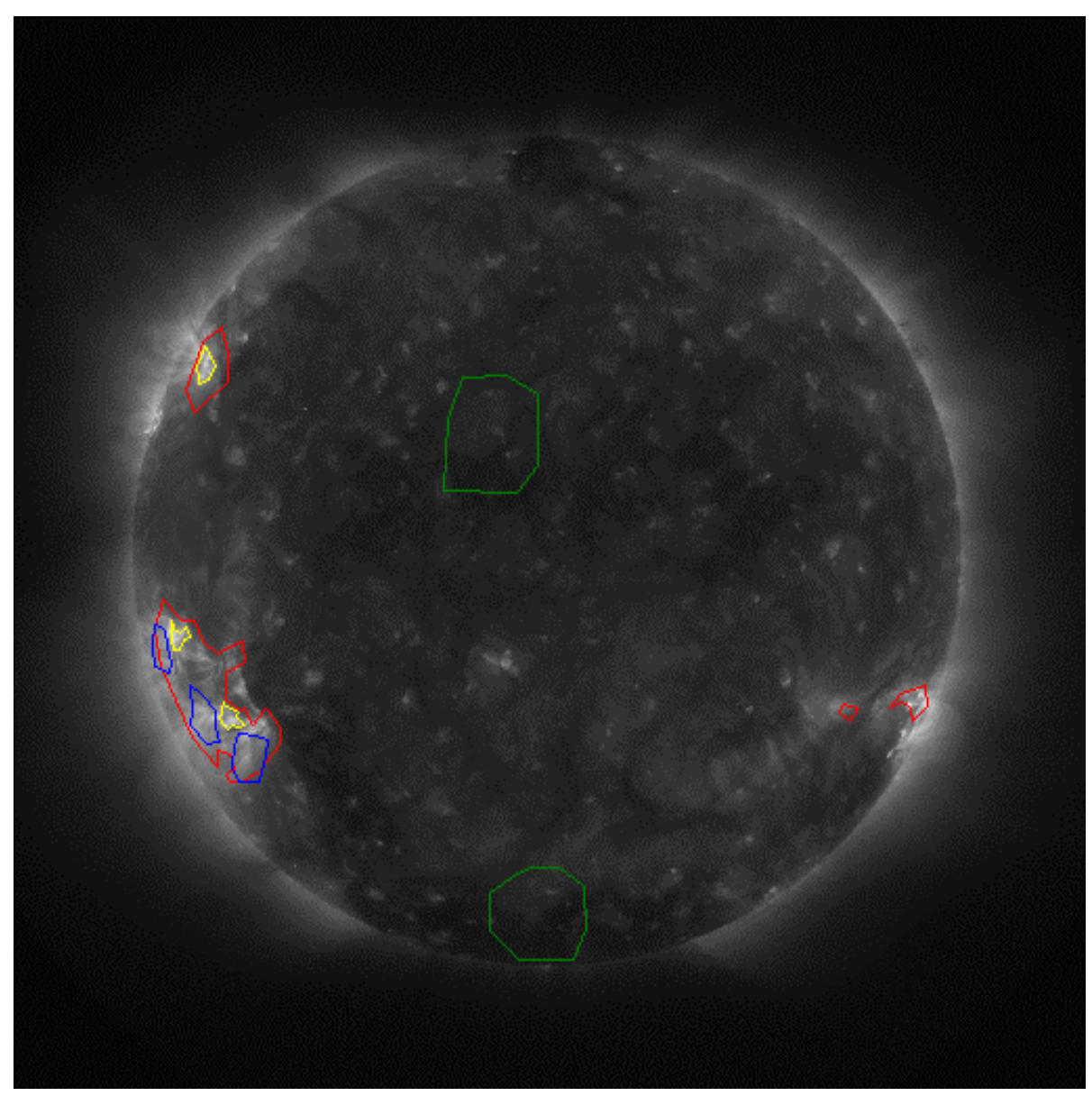

(b) Data 2 (2021/06/24) at 19:27UT

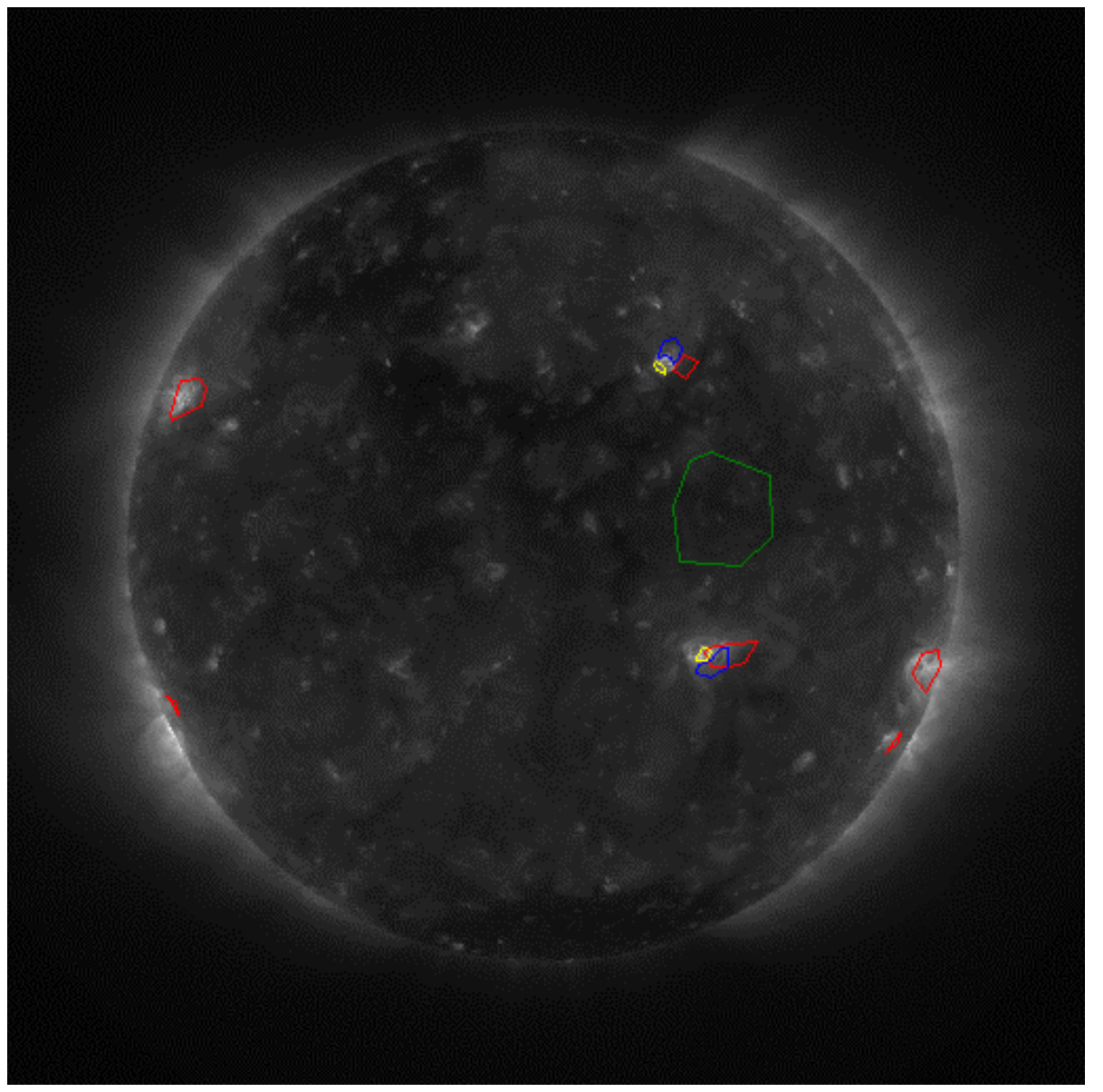

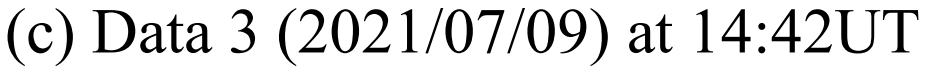

(c) Data 3 (2021/07/09) at 14:42UT

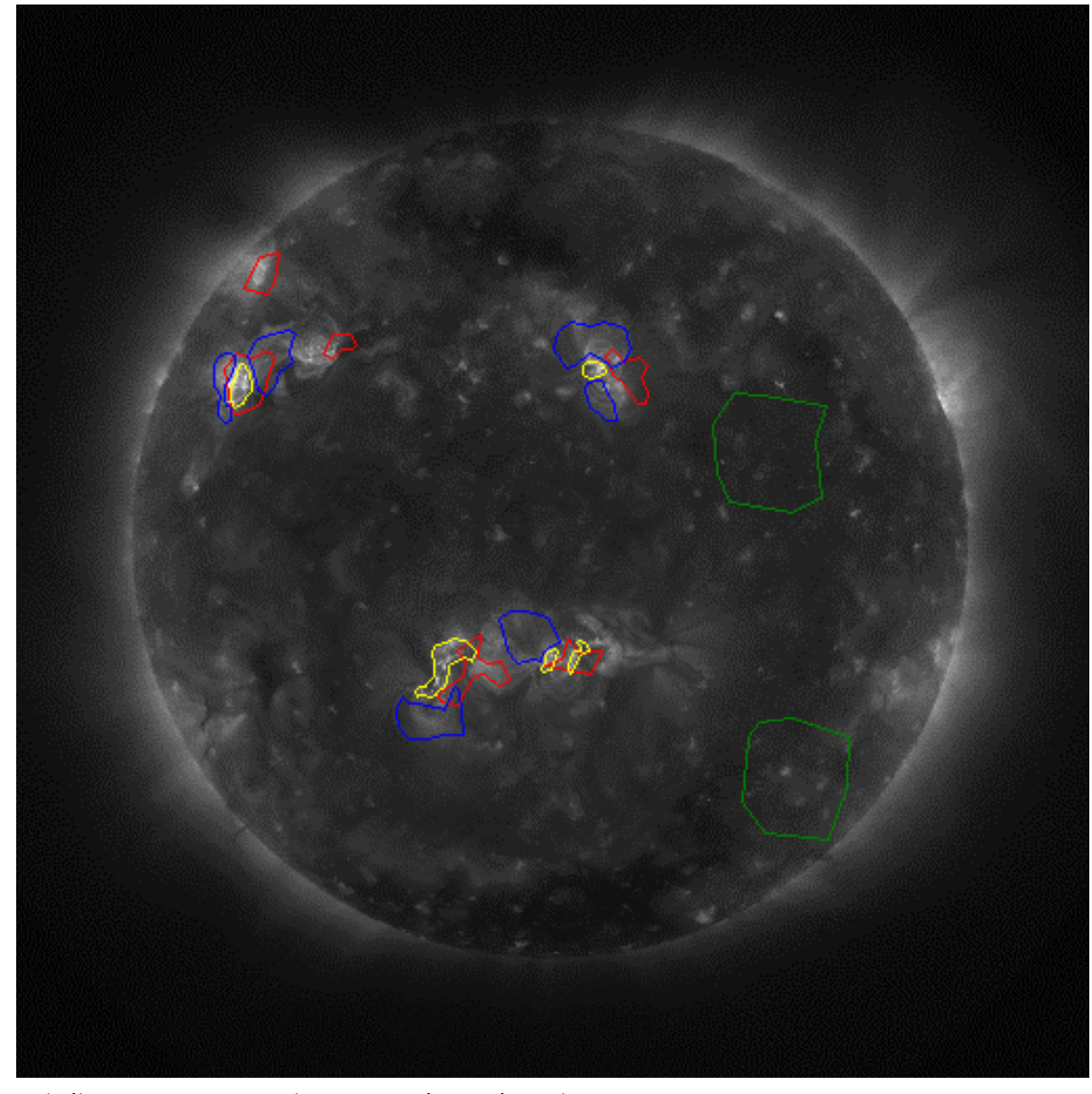

(d) Data 4 (2021/09/05) at 18:00UT

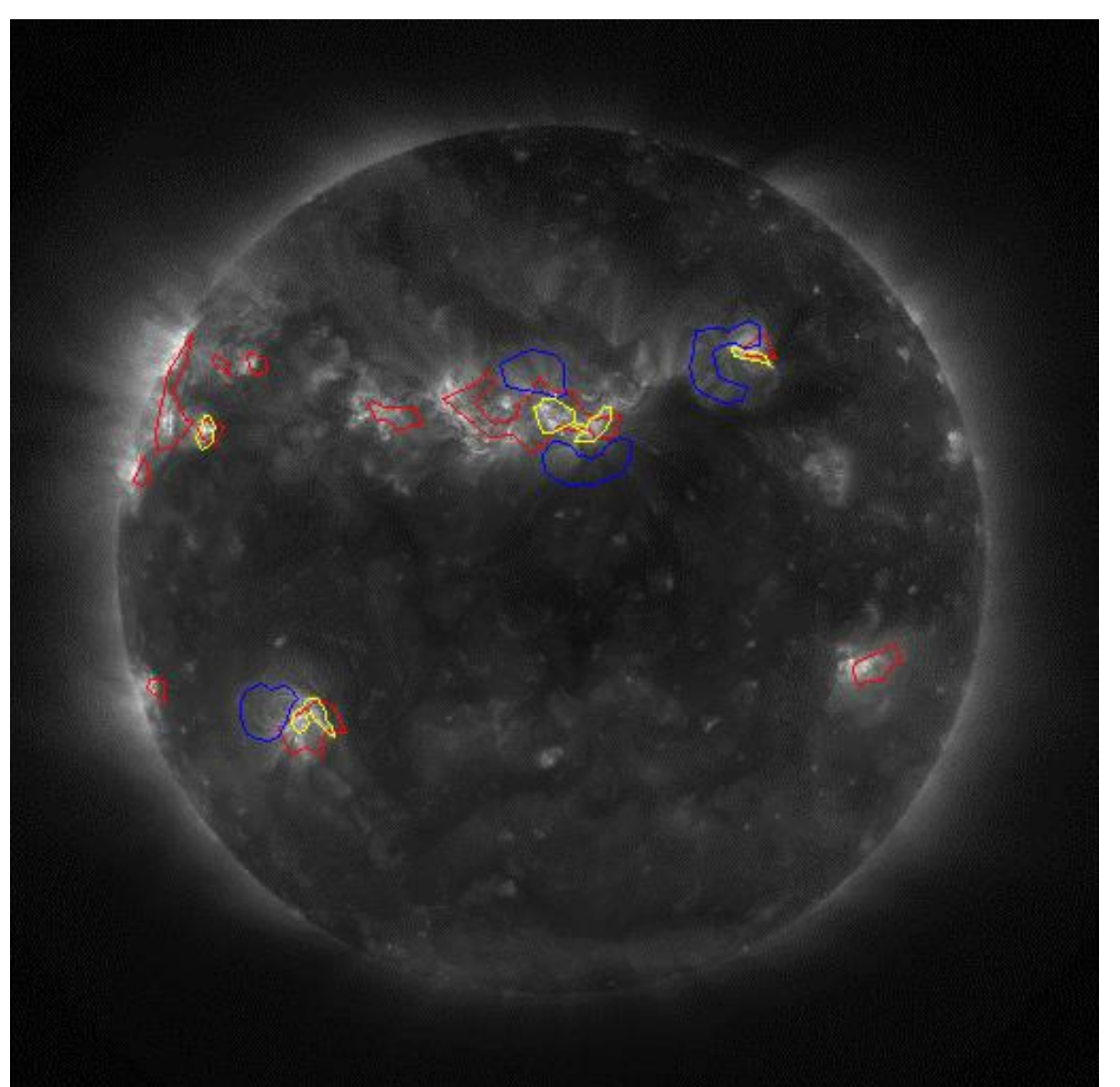

(e) Data 5 (2021/12/19) at 09:16UT

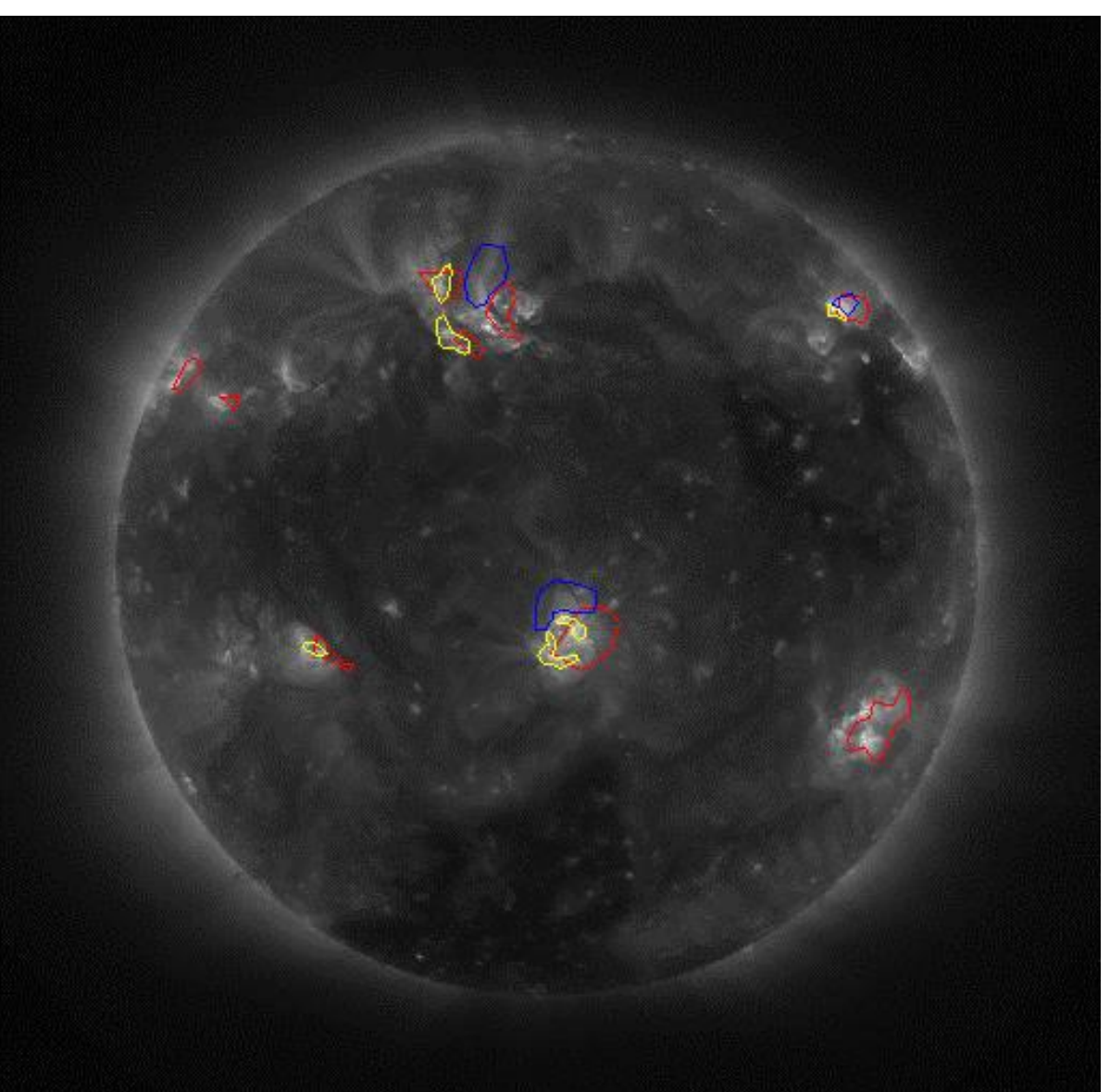

(f) Data 6 (2021/04/24) at 09:01UT

Figure 2 (a-f): SDO AIA images in FITS format showcasing the investigated events: Green labeling denotes areas with no Active Regions (Class 1 – nAR type 1), blue indicates non-flaring Regions outside active region with brightness (Class 2 - nAR type 2), and yellow highlights flaring Active Regions (Class 3 - ARs). The red color represents the ARs captured by HEK through SPoCA model. The illustrated colored areas are used as our training data. Event numbers correspond to those listed in Table 1.

*2.2 Method for 2D Circular Kernel Time Series Transformation of Solar Observatory Images*

The concept of using circular kernel to calculate the 2D Entropy was developed by [41, 42]. Its essence lies in the transformation of a 2D image area using circular kernel of a radius $R$ into a 1-DTS (Figure 3a). The circular kernel of radius $R$ = 14 pixels spans $N$ = 613 pixels, and scans each image with step $S$=8 pixels and generates a dataset of one-dimensional time series. The set of pixels inside the local kernel is converted into 1-DTS data. Next, the algorithm traces the pixel along the connecting line shown in Figure 3b, starting from the center of the kernel ($N$ = 1) and ending with $N$ = 613, leading to the formation of elements of the time series, as shown in Figure 3c. The first element of the series is always equal to the brightness value of the central pixel, relative to which the circular core is located. Areas which are outside the image boundaries (pink color on Figure 3a) are not defined, so the pixel values in these areas are filled by the symmetrical mirroring of the pixels in the image. The number of pixels (*N*) in a circular kernel has a quadratic dependence on the radius [41].

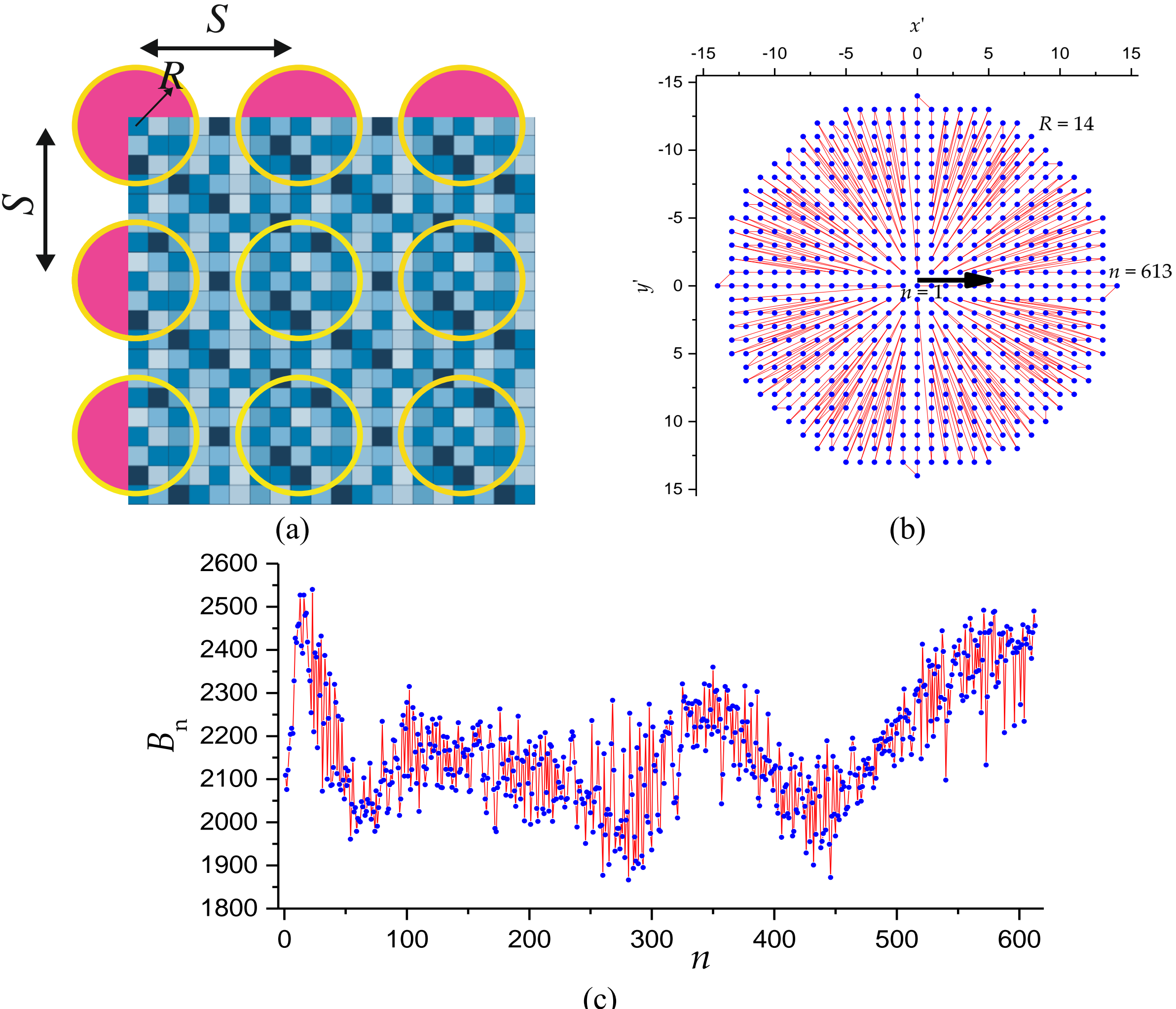

**Figure 3:** Diagrams illustrating the transformation of an image to a one-dimensional time series using a circular kernel: (a) Overview of the transformation process. (b) Detailed view of converting

a two-dimensional pixel distribution into a one-dimensional series for a radius of 14 pixels. (c) Example of a time series generated from a single circular kernel.

### *2.3 The Method for Creating Training Dataset*

Dataset for training a neural network was obtained from 10 SDO AIA observations shown in Table 1. Figures (2 and 4) illustrate the position of the training dataset pixels on the solar observatory images. Areas marked in green correspond to the areas with no Active Regions (class 1; nAR type 1). Areas marked in blue correspond to the areas of non-flaring regions outside active region with brightness (class 2 - nAR type 2), also refers as no active regions. Areas marked in yellow correspond to the flaring Active Regions (class 3-AR). Since one image contains 4096×4096 pixels, to reduce the computational time, the 1-DTS calculation was performed around base pixels, eight pixel apart (see Figure 3a). Thus, 1-DTS was calculated on a 512x512 pixels matrix. The total number of 1-DTS for all the 6 images is 6×512×512= 1 572 864. The dataset used for training the model contained 6606 1-DTS. Each of the three classes contains 2202 1-DTS, so dataset is balanced.

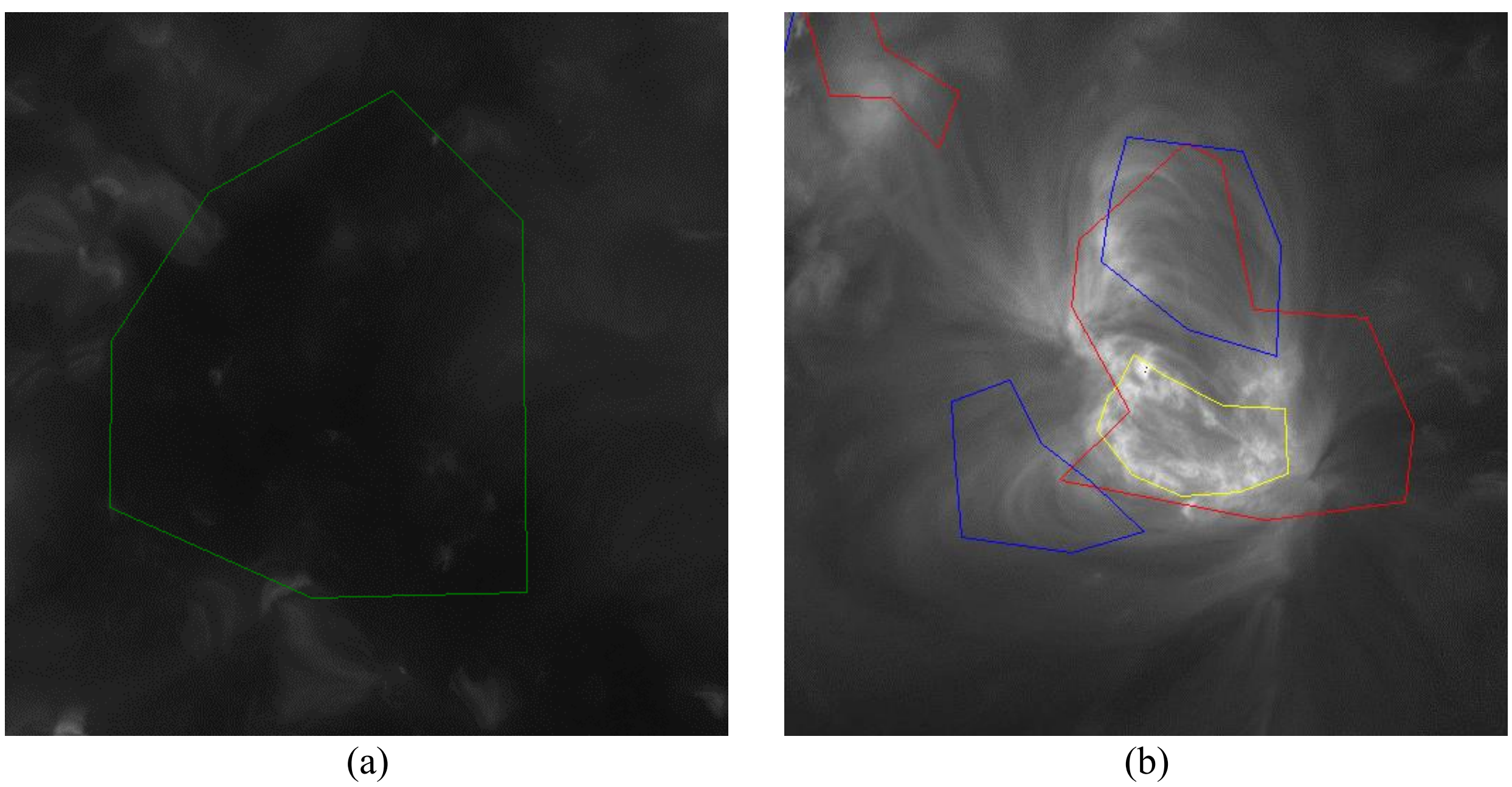

(a) (b)

Figure 4: Axial view of the solar observatory image in FITs format, revealing labels for no Active Regions (nARs, type 1), non-flaring Regions outside active region with brightness (nARs, type 2) and Active Regions (ARs). Green areas indicate nARs (Class 1), blue areas denote nARs outside active region with brightness (Class 2), and yellow areas show flaring ARs (Class 3). Red areas are ARs identified by HEK through the SPoCA model.

### 2.4 Feature selection

**Two different feature selection methods (FSM) were used in our study.**

#### 2.4.1 Feature selection method 1

In FSM 1, we employ statistical and entropy features to extract information from one-dimensional time series (1-DTS) obtained from the SDO AIA observations. For each time series in the training dataset (total 6606 1-DTS ), the Statistical (median value; ($X_{med}$), and 95th percentile; ($X_{95}$)) and Entropy (Distribution Entropy; (DistEn) [50], and Fuzzy Entropy; (FuzzyEn) [51]) characteristics were estimated.

Fuzzy Entropy (FuzzyEn) measures entropy based on fuzzily defined exponential functions for comparison of vectors similarity. It differs from Approximate Entropy and Sample Entropy, which use Heaviside function to calculate the irregularities in a time series data [51]. Fuzzy Entropy is calculated as follows. For a given time series $x(n) = [x(1), x(2), \dots, x(N)]$ with given embedding dimension $(m)$, an $m - vectors$, the Fuzzy Entropy is of the form:

$$X_m(i) = [x(i), x(i+1), \dots, x(i+m-1)] - x0(i) \quad (1)$$

These vectors represent $m$ consecutive $x$ values, starting with $ith$ point, with the baseline $x0(i) = \frac{1}{m}\sum_{j=0}^{m-1} x(i+j)$ removed. Then, the distance between vectors $X_m(i)$ and $X_m(j)$, $d_{ij,m}$ is defined as the maximum absolute difference between their scalar components. Given $n$ and $r$, the degree of similarity $D_{ij,m}$ of the vectors $X_m(i)$ and $X_m(j)$ is calculated using fuzzy function.

$$D_{ij,m} = \mu(d_{ij,m} r) = exp\left(-\frac{(d_{ij,m})^n}{r}\right) \quad (2)$$

The function $\phi_m$ is defined as:

$$\phi_m(n,r) = \frac{1}{N-m}\sum_{i=1}^{N-m}\left(\frac{1}{N-m-1}\sum_{j=1, j\neq i}^{N-m} D_{ij,m}\right) \quad (3)$$

Repeating the same procedure from equation (1-3) for the dimension to $m+1$, vectors $X_{m+1}(i)$ are formed and the function $\phi_{m+1}$ is obtained. Therefore, FuzzyEn can be estimated as:

$$FuzzyEn(m,n,r,N) = \ln \phi_m(n,r) - \ln \phi_{m+1}(n,r) \quad (4)$$

In the computation of Fuzzy Entropy, the embedding dimension $m = 1$ and tolerance r=0.05 x std were used in the ML feature extraction, where std is a standard deviation of $x(n)$.

Distribution Entropy (DistEn) is described based on the distribution of distances between the embedding vectors $X_i$, which are constructed based on the original time series $x_i = [x_1, x_2, \dots, x_N]$ with given embedding dimension $m$. The embedding vectors are formulated as follows:

$$X_i = [x_i, x_{i+1}, \dots, x_{i+m-1}], 1 \leq i \leq N-m \quad (5)$$

For each pair of vectors $X_i$ and $X_j$, the Chebyshev distance $d_{ij}$ is calculated:

$$d_{ij} = \max\{|x_{i+k} - x_{j+k}|, 0 \le k \le m-1\}, 1 \le i,j \le N-m \quad (6)$$

Then, for all distances $d_{ij}$ ($i \neq j$), a distribution histogram is constructed over the entire range of values using $M$ bins. Then, for each bin, the probability $p_t$ ($t$=1..$M$) is calculated based on the frequency approach. Using probability calculations, one can calculate DistEn based on Shannon's definition of entropy as follows (also normalizing the value in the range from 0 to 1):

$$\text{DistEn} = -\frac{1}{\log_2 M}\sum_{t=1}^{M} p_t \cdot \log_2 p_t \quad (7)$$

These Statistical and Entropy measures were applied as features extraction to the 1-DTS obtained from SDO AIA images.

**2.4.2 Feature selection method 2**

In FSM 2, the input to the neural network classifier was directly applied to the one-dimensional time series obtained from the SDO AIA observations. 1-DTS elements were used as features.

2.5 Cross-Validation

To classify the dataset, the Support Vector Classifier (SVC) with radial basis function (RBF) kernel was used, provided by the scikit-learn library [52, 53]. Features' values were standardized by subtracting the mean value of the feature, and dividing it by its standard deviation. Models using one of the two FSMs are further referred to in the text as FSM 1 model and FSM 2 model. The model accuracy assessment consisted of two successive stages. At the first stage, hyperparameters were tuned using the stratified K-fold (K=10) cross-validation [52]. The distribution of classes in each fold approximately corresponds to the distribution in the original dataset. The following hyperparameters were used: the regularization parameter C (in the range from $10^{-2}$ to $10^{5}$) and the kernel parameter gamma (in the range from $10^{-4}$ to $10^{2}$). The best cross-validation accuracy was shown by the model with C=100 and gamma=0.1. To reduce the variance error in assessing the classification accuracy, the second step was used. Using the values of the hyperparameters from the first step, the accuracy of the model was assessed through the method of stratified repeated K-fold cross-validation [54] with N = 10 different partitions into K = 10 folds. The average accuracy value over N = 10 repetitions $A_{KF}$ was subsequently used as an estimate of the accuracy of the classifier. We also calculated the confusion matrix averaged over N=10 repetitions and mean receiver operation curves (ROC) averaged for all (K·N = 100) folds.

2.6 Methods for testing the stability of models to rotational transformation

An important aspect when marking Active Regions is the stability of the result to rotational transformation. When classifying Active Region as shown in Figure 4, the result of applying any method should be rotation invariant. We know that Sunspots and flares can rotate [55] and their marking should not depend on the angle at which they are located relative to the observer. To test the rotation-invariant of our models, we rotated our sampled images 90° clockwise and reclassified the previously identified regions. The match of the result was evaluated by the mismatch index $T_{90^0}$

$$T_{90^0} = \frac{Number\ of\ non-matching\ classifications}{(nAR\ type\ 1)+(nARs\ type\ 2)+AR} \quad (9)$$

The $T_{90^0}$ index is the ratio of the number of non-matching to the total number of ARs and nARs. The closer $T_{90^0}$ approaches 0, the less mismatch there is during rotation. A perfect match corresponds to $T_{90^0}$=0.

2.7 Method of estimating of Generalized Solar Activity (GSA)

We introduce a proxy for estimating the overall solar activity in an observation. This proxy, called the Generalized Solar Activity (GSA) is defined as the ratio of the number of **ARs** pixels to the total number of ARs and nARs.

$$\text{GSA} = \frac{AR}{(nAR\ type\ 1)+(nARs\ type\ 2)+AR} \times 100\% \quad (10)$$

where $AR$ represent the flaring Active Region, $nAR\ type$ 1 is the no Active Regions and $nARs\ type$ 2 is the non-flaring Region outside active region with brightness.

## 3. Results

### *3.1. The Results from Classification of Solar Observatory Images Using Statistical and Entropy Features (FSM 1)*

In Table 2, we present the classification accuracy of the FSM 1 for nAR Type 1, nAR Type 2, and AR instances. The performance is listed in terms of each feature separately (column 1 and 2 of the table), as well as their combined power (column 3 and 4). The best feature is FuzzyEn with classification accuracy $A_{KF}$ of 0.895. The best statistical feature is 95$^{th}$ percentile ($X_{95}$): $A_{KF}$ = 0.873.

Table 2: Classification outcomes for the FSM 1.

| Feature | $A_{KF}$ | Feature | $A_{KF}$ |
|---|---|---|---|
| DistEn | 0.738 | All entropy features | 0.900 |
| FuzzyEn | 0.895 | All statistical features | 0.914 |
| $X_{95}$ | 0.873 | All features | 0.940 |
| $X_{med}$ | 0.840 | | |

The averaged confusion matrix, detailed in Table 3, reveals that the most significant classification errors occur between classes nAR type 1 and nAR type 2, followed by classification errors between nAR and **AR**. However, classification errors between nAR and ARs are notably fewer. Figure 5 shows the histogram of distribution for ARs and nARs (type 1 and type 2) classes in the dataset used for training of the FSM 1. As shown in Figure 5, the distribution of Statistical and Entropy features per class agrees with the conclusions from Table 3, and show that the overlap of the nAR and AR classes according to the features used is minimal. Figure 6 presents the mean ROC curves for each of the 3 classes, performed according to the one-versus-all scheme for the FSM 1.

Table 3: Averaged confusion matrix for the FSM 1 using all features

| | | Predicted labels | | |
|---|---|---|---|---|
| | | nAR type 1 | nAR type 2 | AR |
| Actual labels | nAR type 1 | 2148.4 | 76.5 | 6.1 |
| | nAR type 2 | 47.6 | 1986.6 | 123 |
| | AR | 6 | 138.9 | 2072.9 |

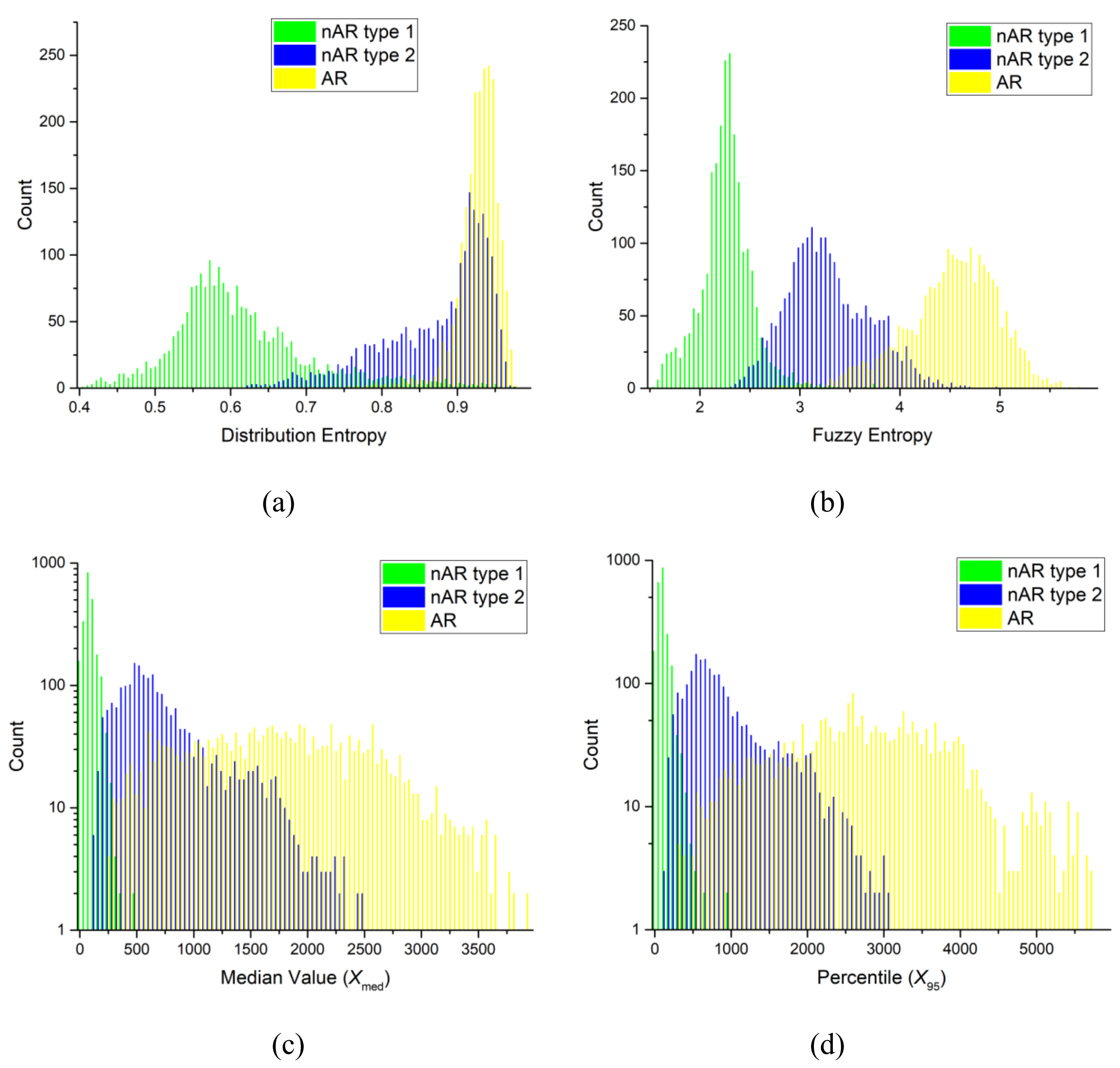

**Figure 5**: Histograms showing the distribution of DistEn, FuzzyEn, Xmed, and X95 values for AR and nAR type 1, and nAR type 2 classes, based on the dataset used for training the FSM 1.

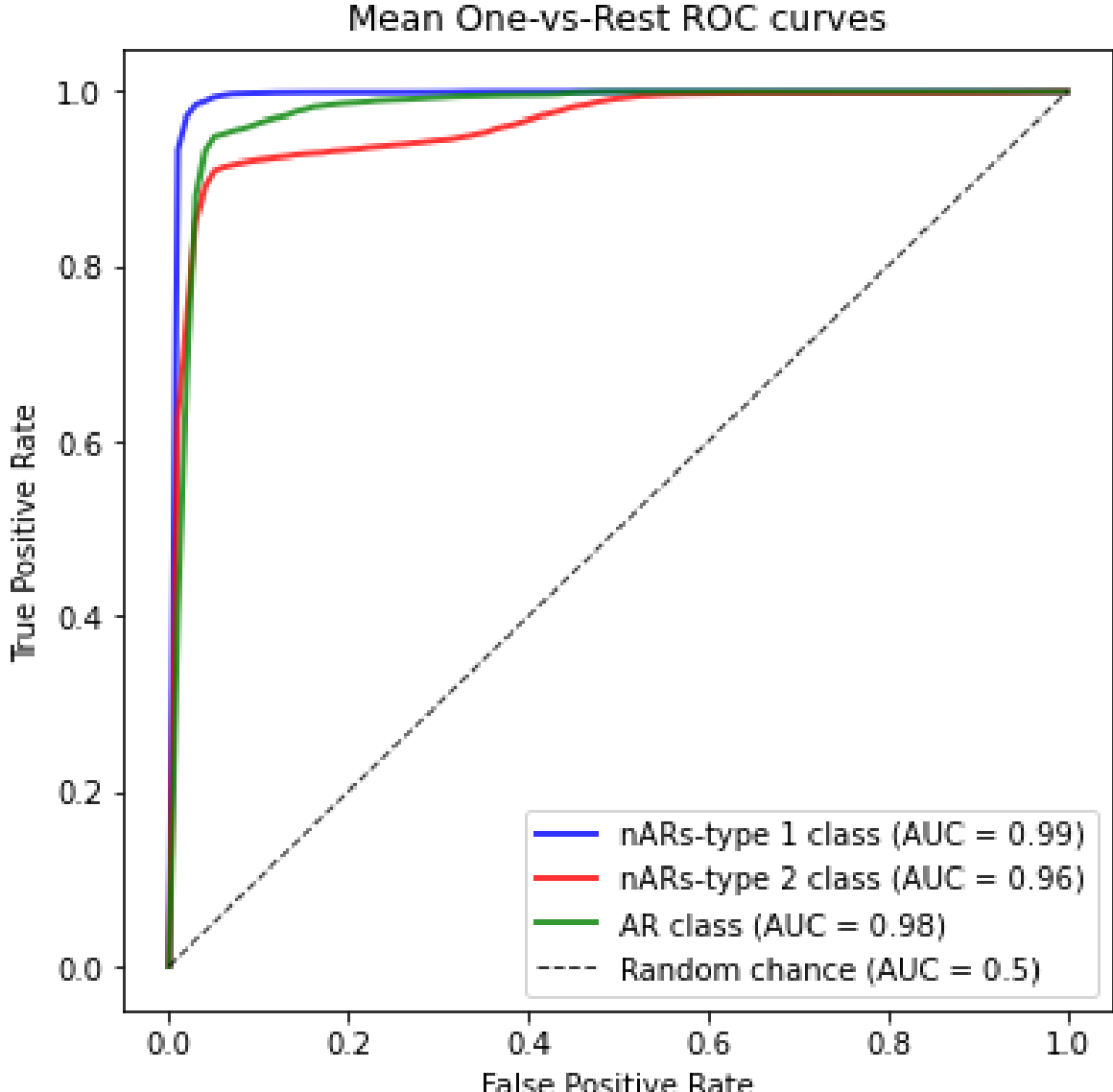

**Figure 6.** Mean ROC curves for nAR type 1, nAR type 2 and AR classes, performed following the one-versus-all scheme for the FSM 1.

The final classification results are shown in Figure 7. FSM 1 reliably identifies the areas of non-flaring Region outside active region with brightness (depicted in blue) and flaring Active Regions (depicted in yellow) in the SDO AIA images investigated.

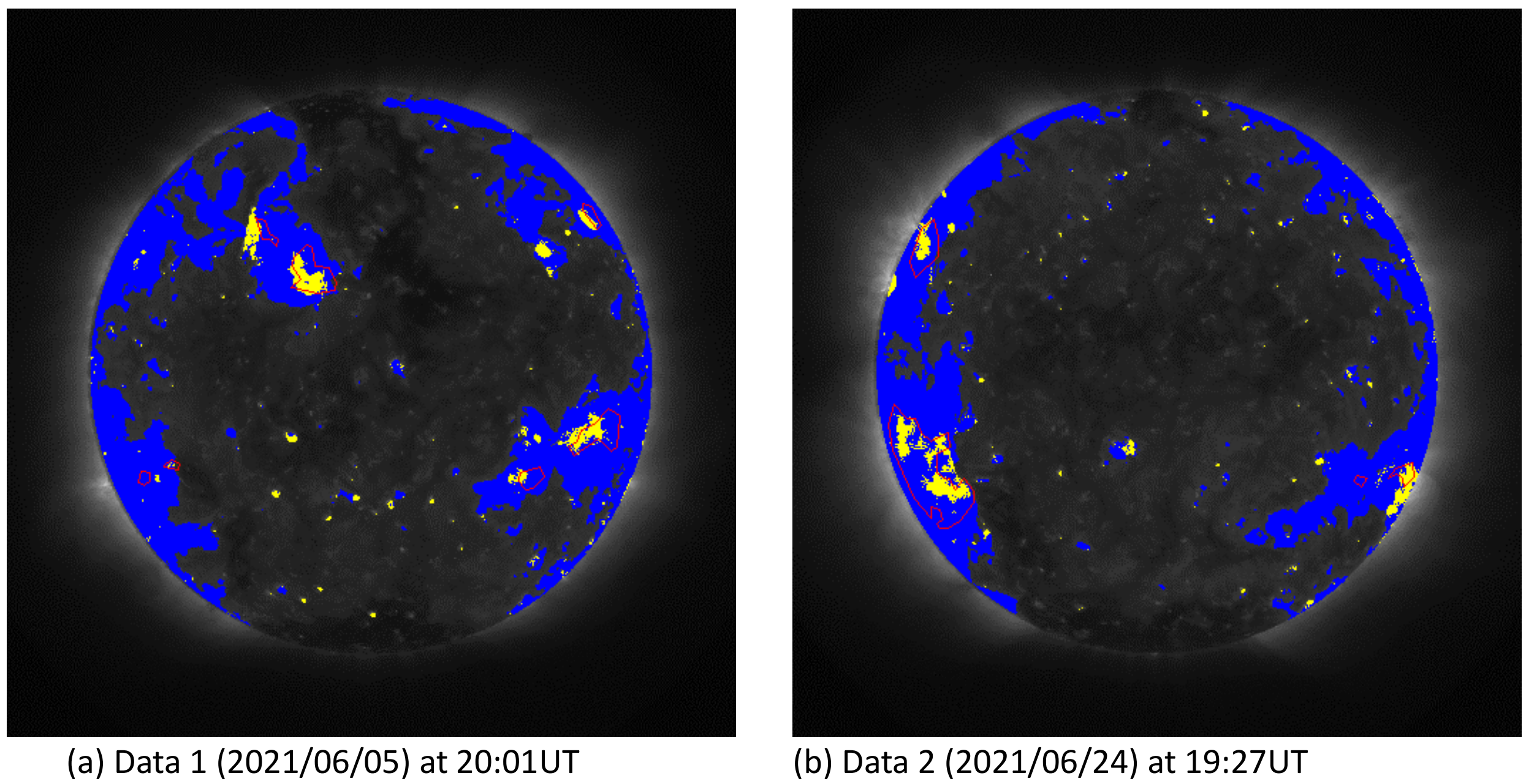

(a) Data 1 (2021/06/05) at 20:01UT

(b) Data 2 (2021/06/24) at 19:27UT

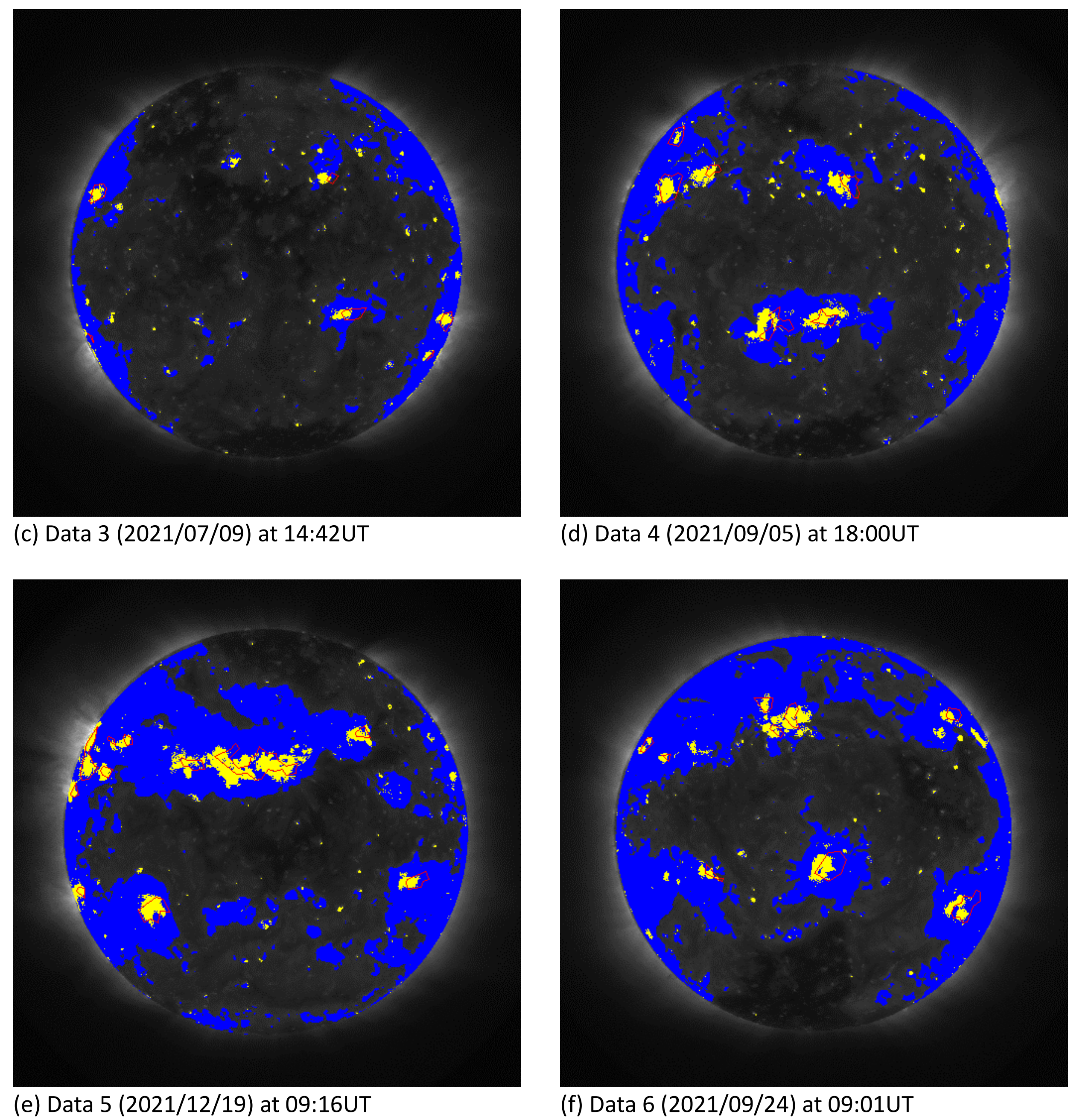

(c) Data 3 (2021/07/09) at 14:42UT

(d) Data 4 (2021/09/05) at 18:00UT

(e) Data 5 (2021/12/19) at 09:16UT

(f) Data 6 (2021/09/24) at 09:01UT

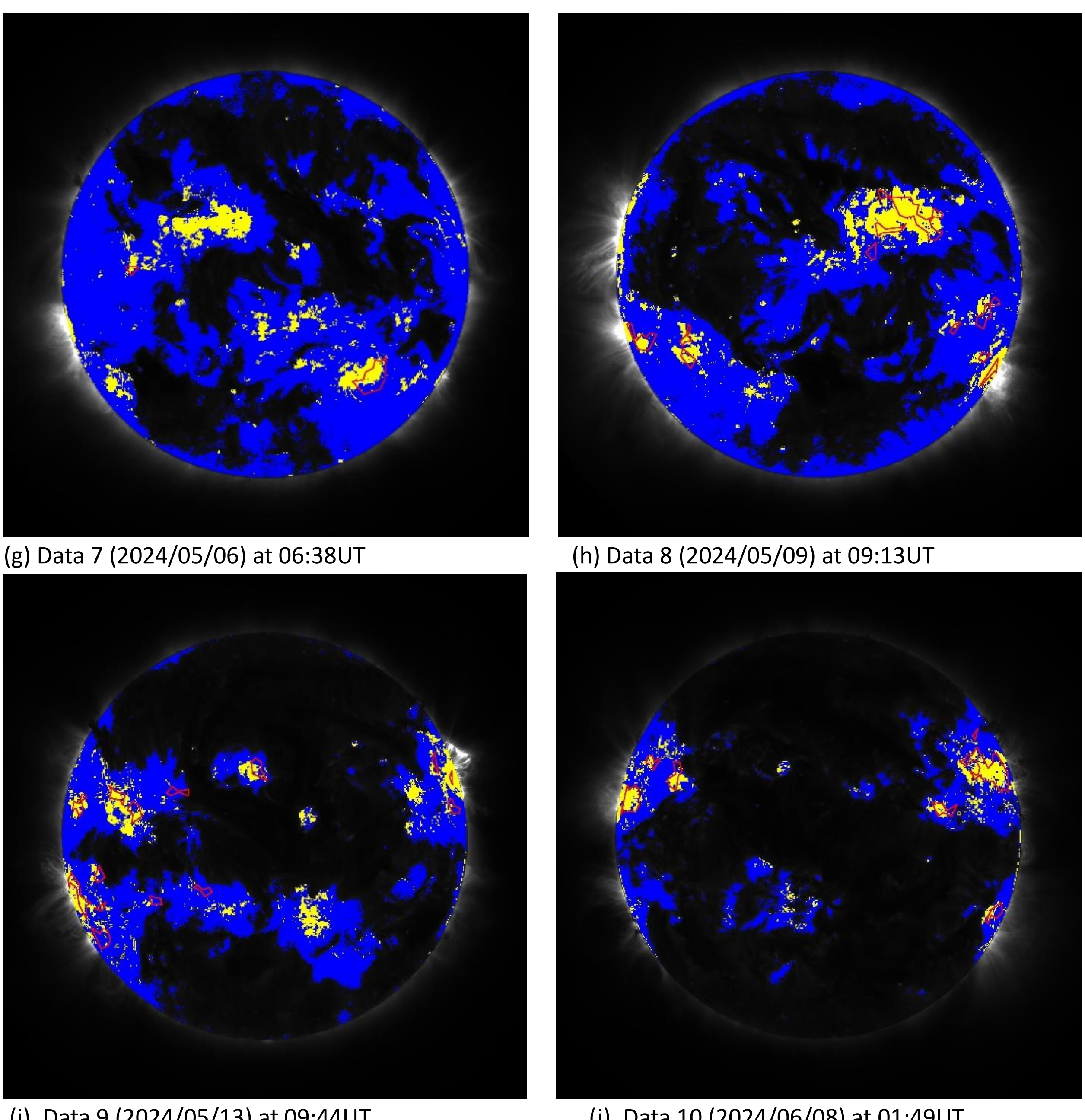

(g) Data 7 (2024/05/06) at 06:38UT

(h) Data 8 (2024/05/09) at 09:13UT

(i). Data 9 (2024/05/13) at 09:44UT

(j). Data 10 (2024/06/08) at 01:49UT

**Figure 7.** The detection of **flaring** ARs in the AIA images using the FSM 1**.** The blue color indicates the captured non-flaring regions outside active region with brightness based on FSM 1 and the yellow color signifies the captured flaring ARs based on FSM 1. The red color is the ARs identified based on HEK through the SPoCA model.

*3.1.1 Robustness of Model to Rotational Transformation (FSM 1)*

The validity checking of the FSM 1 to rotational transformation revealed a mismatch index $T_{90°}\sim$ 0.006 when SDO AIA image "Data 1" of the solar observatory is rotated 90° (see Figure 8).

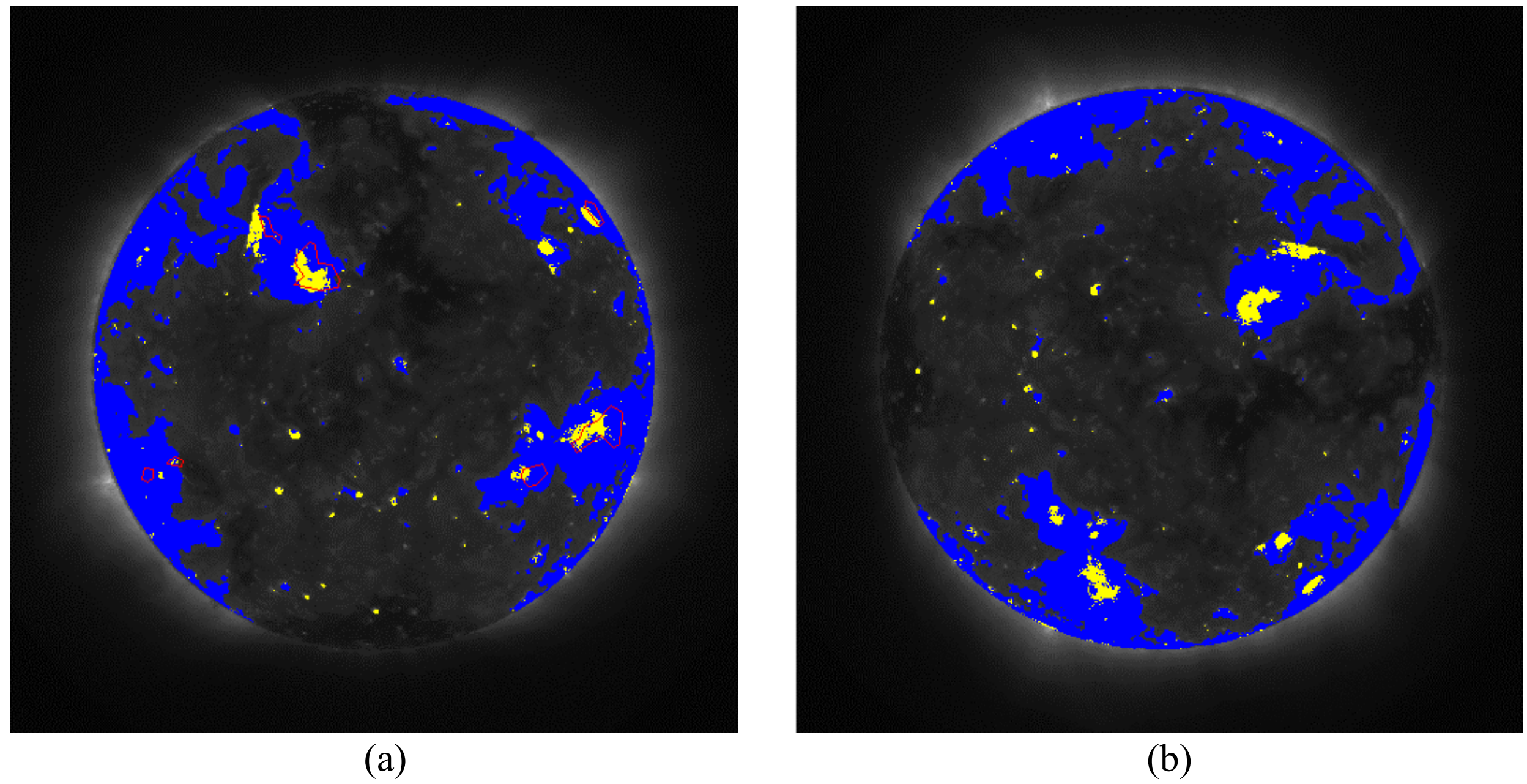

(a) (b)

**Figure 8.** The results of applying the FSM 1 model in image no 1 (2021/06/05 at 20:00UT), after training on the original base before (a) and after rotation by 90 ° (b) ($T_{90}$°~ 0.006).

The low value of the mismatch index signifies that the FSM 1 model is rotation invariant regardless of the rotation invariance of the utilized features. The uniqueness of the proposed spherical kernel lies in the fact that the rotation of the image is equivalent to a cyclic shift of the elements of the time series along it. Figure 9a demonstrates the impact of a rotation by 90° degrees. The gradual rotation of the circular kernel around its axis is equivalent to the displacement of the time series in a sequential, circular order. Figure 9b shows an example of the original 1-DTS shifted by 154 steps, which is equivalent to rotating the image by 90°. The dependence of DistEn and FuzzyEn on the magnitude of the shift in the time series is shown in Figure 9c. The relative deviation (Standard Deviation/Mean) for DisEn is $\sim 1.24 \cdot 10^{-4}$, and for FuzzyEn (Standard Deviation/Mean) is $\sim 6 \cdot 10^{-4}$. Thus, the displacement of the time series has a negligible effect on the entropy characteristics with a variation of about 0.1% of the mean entropy value. Therefore, rotating the entire image, or ARs in the images exhibit rotation-invariance.

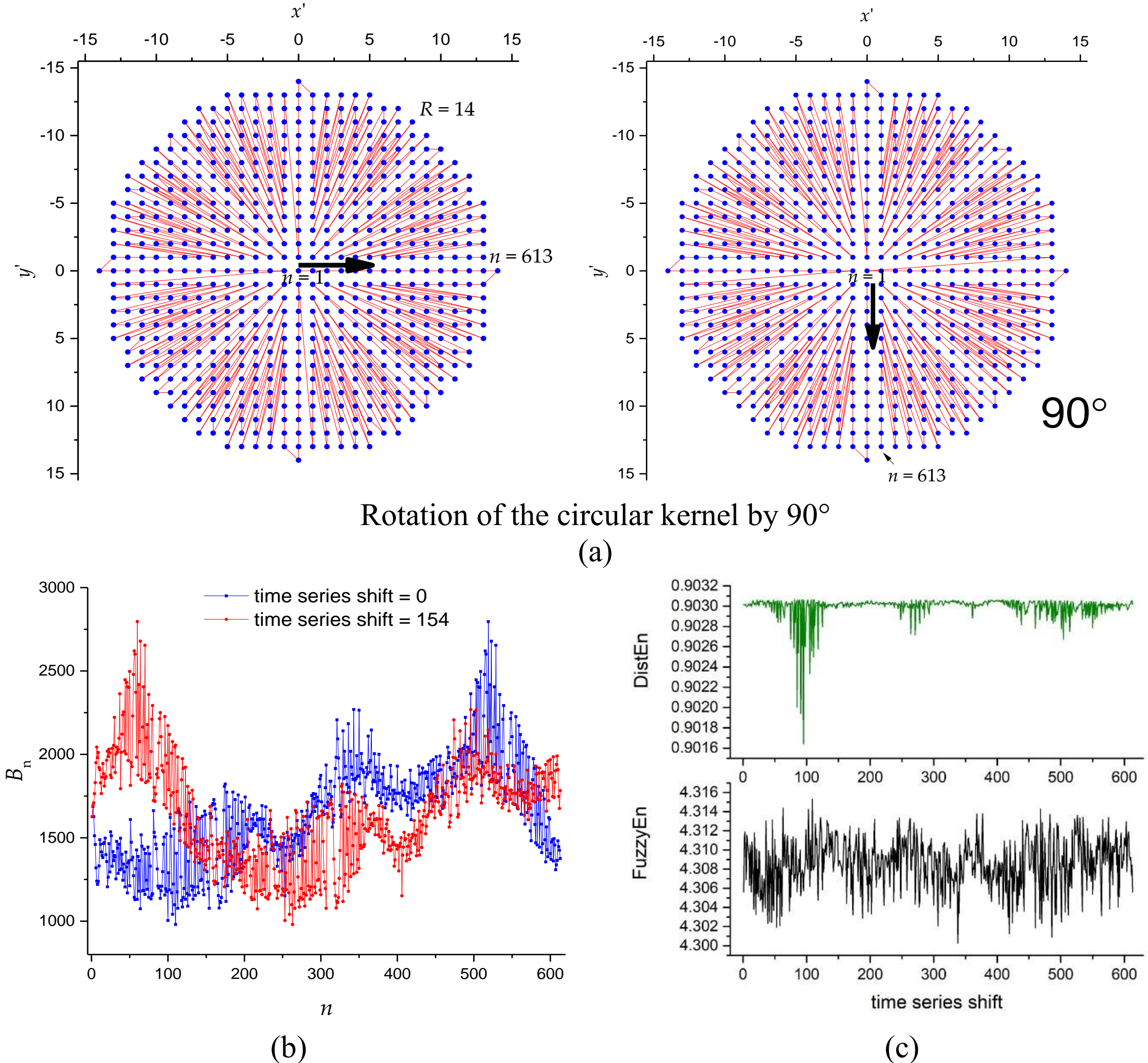

**Figure 9:** (a) Illustration of a 90-degree rotation of a circular kernel, equivalent to a 154-step shift in the time series. (b) The original base image's 1-DTS and its transformation after a 90-degree rotation. (c) Corresponding values of DistEn and FuzzyEn for the time series shift.

### *3.2. The Results from Classification of Solar Observatory Images Using 1-DTS (*FSM 2*)*

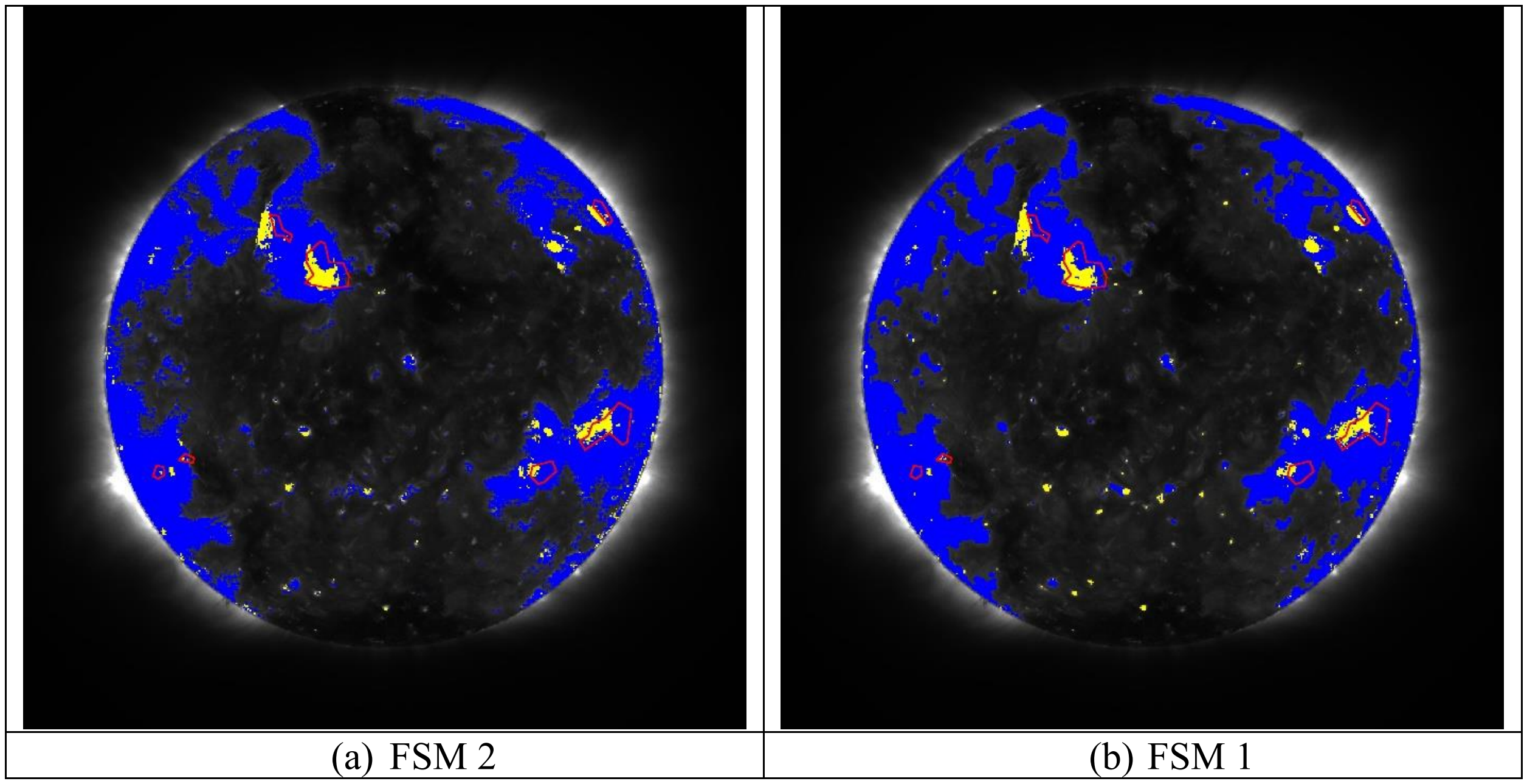

(a) FSM 2 (b) FSM 1

**Figure 10:** The comparison of the flaring AR detection in the SDO AIA image on (2021/06/05) at 20:01UT by FSM 1 and FSM 2. The red color areas are the flaring AR captured by the SPoCA model. The blue color areas represent the non-flaring regions outside active region with brightness (nAR type 2) captured by our proposed model while the yellow color areas are the flaring AR also captured by our model.

FSM 2 detects a similar distribution of flaring ARs and nAR areas in the SDO AIA images as obtained by FSM 1, as shown in Figure 10 (a & b). The computational time for the FSM 2 method using the GSA ver. 1 Python script on an AMD Ryzen 9 3950X 16-Core Processor at 3.49 GHz for one image per thread is approximately 150 seconds. In contrast, the computational time for FSM 1 for image processing per thread is around 9900 seconds. The GSA Python script allows for multithreaded computation, and for 30 threads, the calculation time per image for FSM 1 is reduced to about 5 minutes.

The extended duration required for FSM 1 to process the SDO AIA images is due to the model's need for entropy calculation. Therefore, FSM 2, which involves direct estimation of 1-DTS elements as features, is computationally faster. However, the results obtained from FSM 1 exhibit less noise in SDO AIA imagery information compared to FSM 2 (see Figure 10). Both FSMs capture similar distributions of AR and nAR activities in the SDO AIA images and show nearly similar values of the Generalized Solar Activity (GSA) index.

**Table 4:** Averaged confusion matrix for the FSM 2 model

| | | Predicted labels | | |
|---|---|---|---|---|
| | | nAR type 1 | nAR type 2 | AR |
| Actual labels | nAR type 1 | 2149.6 | 74.1 | 6.6 |
| | nAR type 2 | 50 | 2002.5 | 190.6 |
| | AR | 2.4 | 125.4 | 2004.8 |

The comparison of the averaged confusion matrix (Table 4) for the two models reveals that the FSM 2 model makes more errors in determining the AR class and fewer errors in determining nAR type 2. Specifically, FSM 2 has more misclassifications of AR as nAR type 2 and fewer misclassifications of nAR type 2 as AR. This indicates that the FSM 2 model is more likely to classify an image pixel as nAR type 2 rather than AR if the test sample is similar to training samples from both classes.

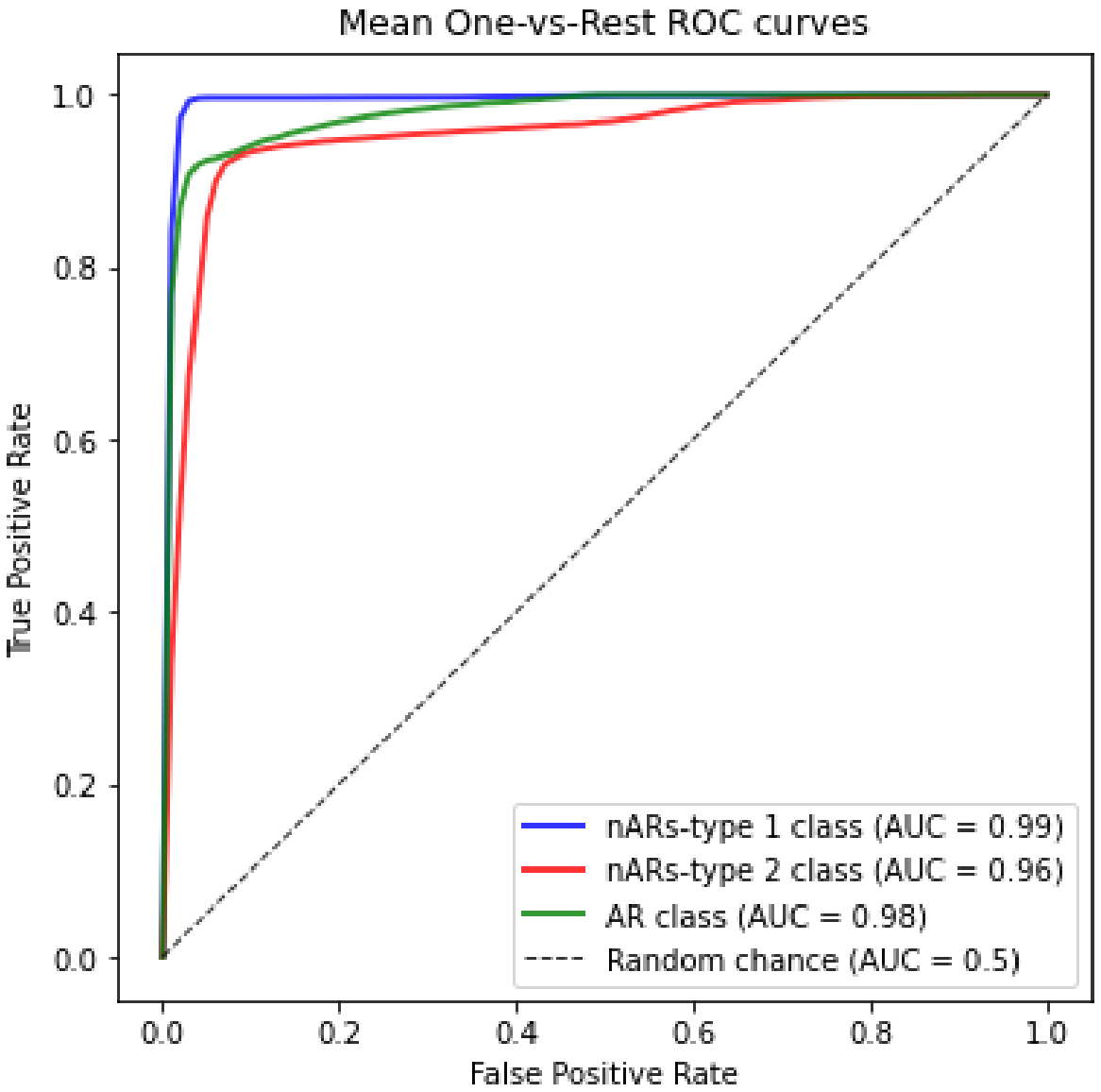

Figure 11. Mean ROC curves for nAR type 1, nAR type 2 and AR classes, performed following the one-versus-all scheme for the FSM 2 model.

**Figure** 11 presents the mean ROC curves for each of the 3 classes, performed according to the one-versus-all scheme for the FSM 2 model. The AUC values for all three curves of both models correspond to each other, but the shape of the curves for AR and nAR type 2 is slightly different.

*3.2.1 Robustness of Model to Rotational Transformation (FSM 2 model)*

The validation of FSM 2 for rotational transformations revealed a mismatch index $T_{90^\circ} \sim 0.06$ when rotating the SDO AIA image 'Data 1' from the solar observatory by 90°. This value is an order of magnitude higher compared to FSM 1; however, it remains low, indicating the robustness of FSM 2 to rotation, as it uses the circular kernel proposed in this work. FSM 1 exhibits higher rotational stability due to its utilization of statistical and entropy features, contrasting with the higher error introduced when using 1-DTS as input features in FSM 2.

3.3 The Results from the Generalized Solar Activity (GSA)

We display in Table 4, the values of GSA for the solar flare events investigated in this study. The GSA values comprising of results estimated from the two FSM models. The FSM 1 model estimates GSA in the SDO image through the concept of entropy measures while the FSM 2 model estimate GSA in the SDO AIA image directly. In the table, the GSA values revealed that on May 9, 2024 at 09:13UT tagged image Data 8 depicts the highest values of activity with GSA equal to 5.62% and 5.51% for FSM 1 and FSM 2 model respectively. The images Data 7, 9, 10, and 5 corresponding to May 6th, 2024 at 06:18UT; May 13th at 09:44UT; June 8th, 2024 at 01:49UT and December 19th, 2021 at 09:16UT also depicts higher values of GSA (4.77%, 4.59%; 4.90%, 3.87%; 3.26%, 2.51% and 2.33%, 2.27%) for both FSM 1 model and FSM 2 model. The high GSA values obtained in the SDO AIA images investigated signifies high activities of flaring events during this period. The least degree of solar activity depicting lowest value of GSA obtained by our proposed method is found to be associated with the image Data 3 on July 9th, 2021 at 14:42UT. Notably, this observation depicts a strong agreement between the GSA values and the solar flaring activities in the SDO AIA image, which can be visually verified in Figure 7, where image Data 3 reveals a minimal solar activity compared to other image Data investigated.

**Table 4.** The values of characteristics of generalized solar activity (GSA) for the selected events used in this study

| S/N | GSA FSM 1 model | GSA FSM 2 model |
|---|---|---|
| Data 1 (2021/06/05) | 1.36% | 1.30% |
| Data 2 (2021/06/24) | 1.47% | 1.51% |
| Data 3 (2021/07/09) | 1.01% | 0.99% |
| Data 4 (2021/09/05) | 1.13% | 1.20% |
| Data 5 (2021/12/19) | 3.81% | 3.71% |
| Data 6 (2021/09/24) | 2.33% | 2.27% |
| Data 7 (2024/05/06) | 4.77% | 4.59% |
| Data 8 (2024/05/09) | 5.62% | 5.51% |
| Data 9 (2024/05/13) | 4.90% | 3.87% |
| Data 10 (2024/06/08) | 3.26% | 2.51% |

3.5 Comparism of the proposed model and the HEK SPoCA model

The red regions in the observation shown in Figures 7 and 10a (see supplementary material for other SDO AIA images results obtained from FSM 2) depicts the flaring ARs activities captured by the HEK SPoCA model. The blue color is the non-flaring regions outside active region with brightness (nAR-type 2) also known as no active regions captured by our models and the yellow color represents the flaring ARs captured by the same model. It was noticed that the SPoCA model did not capture all the flaring ARs activities in the presented observations. For instance, from Figure 7(c-j) it was shown that the SPoCA model did not identify all the ARs activities. However, our model initiated through the concept of classifying the ARs activities into no Active Region (nARs-type 1), non-flaring region outside active region with brightness (nAR-type 2) and flaring ARs (AR) addresses this shortcoming. In that not all the yellow colors captured by our model revealing mini-flaring ARs in the SDO AIA image were also captured by HEK SPoCA model. Notably, SPoCA model only capture few areas of this flaring ARs activities. Our model identifies the areas of flaring ARs and also captures the mini-flaring ARs areas in the SDO AIA images in yellow color. Observably, the identification of ARs activities from SDO AIA images by our model is noticed to be associated with areas of eruption of plasma. Furthermore, we also noticed that our model captures even the mini-flaring active regions which the SPoCA model was unable to capture.

## 4. Discussion

The feature extraction from the 1-DTS of the solar observatory images using Statistical measures (median Value, $X_{med}$, and 95th percentile, $X_{95}$) gives high value of classification accuracy. However, the classification accuracy for 95th percentile $X_{95}$ reveals higher value compared to median value ($X_{med}$). These Statistical measures ($X_{med}$) and ($X_{95}$) reveal potential usefulness of the tools to extract information from the 1-DTS obtained from SDO AIA images. The results of classification accuracy for the Entropy measures demonstrate that FuzzyEn depicts the highest value of classification accuracy compared to DisEn. This reveals the potency of the FuzzyEn to sensitively capture more information regarding ARs activities from the 1-DTS obtained from the SDO AIA images when compared with DisEn. The feature extraction using the Statistical features also provides discriminative information regarding ARs and nARs in the observations. Notably, the distribution for DisEn, FuzzyEn, $X_{med}$ and $X_{95}$ further, corresponding to AR and nAR classes confirms the value of the Entropy and Statistical measures used in this study. Both the Statistical and Entropy measures are useful diagnostics tools in capturing of ARs from SDO AIA images. Interestingly, the FSMs model accurately captures the areas of the images associated with flaring ARs. In the areas of the AIA image associated with nARs, the model also highlights activities of non-flaring Region outside active region with brightness as shown in Figure 7. This is because our model is designed to capture flaring ARs and non-flaring active region outside active region with high brightness.

Calculating Fuzzy Entropy for a large number of series takes quite a long time. Since we calculate a 512x512 pixels matrix, we get a fairly precise image of the active areas. To calculate on one thread, the time for one image takes about 16 minutes. The technology itself makes it possible to parallelize this calculation and its adaptation on graphics accelerators is a topic for further work.

In addition, there is an idea of approximating entropy using a neural network, for example, a perceptron, see our paper [56]. This approach can lead to faster computations. All these works can be the topic of further research.

The feature extraction obtained from the 1-DTS of the SDO AIA image subjected to the rotational transformation test for the FSM 1 model is shown in Figure 8. The validation of feature extraction for rotational transformation revealed that the FSM 1 model provides high classification accuracy. After rotating the SDO AIA image by 90°, there was no noticeable difference, with $T_{90°} \sim 0.006$, suggesting that the model is rotationally invariant.

Furthermore, the rotational transformation for FSM 2 revealed a mismatch index $T_{90°} \sim 0.06$, also indicating that the FSM 2 model is rotationally invariant. A generalizing characteristic of AR activities (GSA) in the SDO AIA images estimated for both FSM 1 and FSM 2 shows that the higher the AR activities in the SDO AIA images, the higher the value of GSA. This observation of GSA estimation can serve as an index to determine general solar activity at a particular moment.

Both FSM 1 and FSM 2 models capture similar results of AR and nAR activities in the SDO AIA images investigated (see supplementary materials). However, the computational speed of FSM 2 is faster compared to FSM 1.

In comparing of the HEK SPoCA ARs model and our method. The article regarding SPoCA model done by [2] estimated the filling factor of all pixels in the SDO AIA images from March 1, 1997 to August 17, 2011 belonging to AR, Quite Sun (QS) and Coronal Hole (CH) respectively, that generalize the Sun activity from March 1, 1997 to August 17, 2011.The filling factor involves studying the evolution of AR, QS, or CH properties over the solar cycle by estimating the total intensity of a region through the sum of all pixel values of the pixels inside the region of AIA image. According to [2] filling factor of a region is defined as the raw area of a region in unit such that the filling factor of the total solar disk equals to 1. We noticed a similar observation in our present study where we estimated the generalized solar activity (GSA) of ARs activities through our method. Our approach gives a more precise GSA value in that, the activities of the flaring ARs is estimated for each SDO AIA images investigated. Furthermore, the detection of the ARs and nARs in SDO AIA image based on our proposed method captures all the ARs and nARs precisely compared to SPoCA model. The comparison of these models is shown in Figure 8. The red circle corresponds to the ARs detected by SPoCA model, while the blues colors correspond to areas of non-flaring region outside active region with brightness and the yellow colors are the flaring ARs detected by our proposed method. It is seen that our approach captures even mini-flaring active regions and flaring active regions unveiling plasma eruption. The SPoCA model did not take into account the mini-flaring active regions in the SDO AIA images investigated. Our further work is to measure precise dynamics of the Sun activity in long period with small time step and make available the python code for the scientific community. Our present study is to introduce the concept of our approach and its potency in capturing the Sun activity of ARs. Also, Convolutional neural networks are very popular and have been used in many image segmentation applications. However, in this work, we intend to unveil the full potentials of Entropy features and proposed a new method based on converting a circular kernel into a one-dimensional time series. We hope that this method can serve as a complement to the CNN method, and its joint use can become a promising direction for further research.

## 5 Conclusion

This study has developed a novel method for tracking Active Regions (ARs) in AIA images from the solar observatory. The images were transformed into one-dimensional time series using a 2D circular kernel. Both statistical and entropy measures were employed as a feature selection method (FSM 1), and the 1-DTS elements were used as features for FSM 2 to effectively identify and analyze the flaring ARs and non-flaring regions within the SDO AIA images. The FSM models achieved consistent classification scores, indicating the robust nature of the system against variations in image orientation. This robustness is exemplified by the model's stability in classification accuracy, requiring fewer training datasets to achieve a $T_{90^\circ}$ of 0.006 for FSM 1 and $T_{90}$ ~0.06 for FSM 2. Notably, Fuzzy Entropy demonstrated superior classification accuracy compared to distribution entropy, underscoring its effectiveness. Both Statistical measures, $X_{med}$ and $X_{95}$, also indicated high classification accuracies. The Entropy measures proved to be exceptionally effective, suggesting that they are powerful tools for feature extraction in detecting ARs in SDO AIA imagery. This approach not only captures ARs but also non-active regions (nARs) of solar activity, illustrating its comprehensive applicability. Furthermore, we established a link between the level of ARs activity and a generalized characteristic of ARs activities (GSA) in the SDO AIA images. Our findings suggest that increased ARs activity **agrees** directly with higher GSA values, potentially offering new insights into solar dynamics and its impacts on space weather forecasting. This correlation could be crucial for improving the accuracy of predictions and understanding the sun's influence on Earth's environment.

**Author Contributions:** Conceptualization, I.A.O and A.V.; methodology, I.A.O, A.V., and M.B. A.A; software, A.V. and M.B.; validation, I.A.O., A.V., M. B., and A.A., O.I.O; formal analysis, I.A.O., A.V., and A.A, M.B; investigation, I.A.O and A.V; resources, A.V. and I.A.O; data curation, I.A.O., and A.V; writing—original draft preparation, I.A.O., A.V., M.B., and I.O.O; writing—review and editing, I.A.O., A.V., M.B., and O.I.O; visualization, I.A.O., and A.A, A.V.; supervision, I.A.O, and A.V.; project administration, I.A.O and A.V.; funding acquisition, A.V.

**Funding:** This research was supported by the Russian Science Foundation (grant no. 22-11-00055, https://rscf.ru/en/project/22-11-00055/, accessed on 30 March 2023).

**Data Availability Statement:** The Python code for our proposed method, Solar Active Detection via a 2D Circular Kernel with the SDO AIA images used in this study, is available in the supplementary materials. Additionally, the solar observatory images used in this study can be accessed through the archive of the Joint Science Operations Centre (JSOC) Database (soc.stanford.edu/AIA/AIA_jsoc.html).

**Acknowledgments:** The authors express their gratitude to Azim Ahmadzadeh for valuable comments made in the course of the article's translation and revision. Special thanks to the editors of the journal and to the anonymous reviewers for their constructive criticism and improvement suggestions. Finally, Dr. Oludehinwa express his appreciation to the NASA Living With A Star 2020 Heliosphere Summer School, been among the beneficiary of the program gave an insight that lead to this project. The authors would like to acknowledge the Joint Science Operations Centre (JOSC) for making the solar observatory images available for research purpose.

**Conflicts of Interest:** The authors declare no conflict of interest

**Supplementary Materials:** the python code of our proposed method and the SDO AIA images used in this study with results can be access from the supplementary material.